\documentclass[prb,twocolumn,superscriptaddress,floatfix,showkeys]{revtex4}
\usepackage{graphicx}
\usepackage{amsmath,subfigure,epsfig,psfrag}
\usepackage{comment}
\usepackage{commath}
\usepackage{graphicx}
\usepackage{amssymb}
\usepackage{xcolor}
\usepackage{soul}
\usepackage{listings}
\usepackage{rotating}
\usepackage{latexsym}
\usepackage{amsfonts}
\usepackage{amssymb}
\usepackage{wasysym}
\usepackage{xfrac}
\usepackage{mathtools}
\usepackage{multirow}
\makeatletter
\renewcommand*\env@matrix[1][\arraystretch]{%
  \edef\arraystretch{#1}%
  \hskip -\arraycolsep
  \let\@ifnextchar\new@ifnextchar
  \array{*\c@MaxMatrixCols c}}
\makeatother
\newcommand{\SUMP}[2]{\relax\ensuremath{\sum\limits_{#1}^{#2}\hspace{-1.5em}{\phantom{\sum}}^\prime}}
\begin{document}

\title{Perturbation Theory Treatment of Spin-Orbit Coupling. III: Coupled Perturbed Method for Solids}

\author{Jacques K. Desmarais}
\email{jacqueskontak.desmarais@unito.it}
\affiliation{Dipartimento di Chimica, Universit\`{a} di Torino, via Giuria 5, 10125 Torino, Italy}

\author{Alberto Boccuni}
\affiliation{Dipartimento di Chimica, Universit\`{a} di Torino, via Giuria 5, 10125 Torino, Italy}

\author{Jean-Pierre Flament}
\affiliation{Universit{\'e} de Lille, CNRS, UMR 8523 --- PhLAM --- Physique des Lasers, Atomes et Mol{\'e}cules, 59000 Lille, France}

\author{Bernard Kirtman}
\affiliation{Department of Chemistry and Biochemistry, University of California,
Santa Barbara, California 93106, USA}

\author{Alessandro Erba}
\email{alessandro.erba@unito.it}
\affiliation{Dipartimento di Chimica, Universit\`{a} di Torino, via Giuria 5, 10125 Torino, Italy}

\date{\today}
\begin{abstract}
A previously proposed non-canonical coupled-perturbed Kohn-Sham density functional theory (KS-DFT)/Hartree-Fock (HF) treatment for spin-orbit coupling is here generalized to infinite periodic systems. The scalar-relativistic periodic KS-DFT/HF solution, obtained with a relativistic effective core potential, is taken as the zeroth-order approximation. Explicit expressions are given for the total energy through 3rd-order, which satisfy the 2N + 1 rule (i.e. requiring only the 1st-order perturbed wave function for determining the energy through 3rd-order). Expressions for additional 2nd-order corrections to the perturbed wave function (as well as related one-electron properties) are worked out at the uncoupled-perturbed level of theory. The approach is implemented in the \textsc{Crystal} program and validated with calculations of the total energy, electronic band structure, and density variables of spin-current DFT on the tungsten dichalcogenide hexagonal bilayer series (i.e. WSe$_2$, WTe$_2$, WPo$_2$, WLv$_2$), including 6p and 7p elements as a stress test. The computed properties through second- or third-order match well with those from reference two-component self-consistent field (2c-SCF) calculations. For total energies, $E^{(3)}$ was found to consistently improve the agreement against the 2c-SCF reference values. For electronic band structures, visible differences w.r.t. 2c-SCF remained through second-order in only the single-most difficult case of WLv$_2$. As for density variables of spin-current DFT, the perturbed electron density, being vanishing in first-order, is the most challenging for the perturbation theory approach. The visible differences in the electron densities are, however, largest close to the core region of atoms and smaller in the valence region. Perturbed spin-current densities, on the other hand, are well reproduced in all tested cases.
\end{abstract}

\maketitle

\section{Introduction}

In modern electronic structure programs, relativistic effects are typically accounted for through self-consistent field (SCF) treatments, either in a two- or four-component spinor basis (2c-SCF or 4c-SCF),\cite{chang1986regular,buenker1984matrix,almlof1985variational,douglas1974quantum,dyall1997interfacing,iliavs2005theoretical,kutzelnigg2006quasirelativistic,liu2007quasirelativistic,liu2006infinite,liu2009exact,peng2007making,sikkema2009molecular,seino2008examination,seino2012local,saue1999quaternion,jo1996relativistic} or by perturbation methods.\cite{rutkowski1986relativistici,rutkowski1986relativisticii,rutkowski1986relativisticiii,kutzelnigg1996stationary,van1995relativistic,stopkowicz2011direct,stopkowicz2019one,cheng2013spin,cheng2018perturbative,desmarais2021perturbationI,desmarais2021perturbationII} 
Within these approaches, a particularly convenient representation of the Dirac equation from a computational perspective is provided through the relativistic effective core potential (RECP), including both scalar-relativistic and spin-orbit coupling (SOC), or non-scalar, effects.\cite{dolg2012relativistic,ermler1981ab} In Part II of this series, we developed a molecular pure-state coupled perturbed Kohn-Sham (CPKS) density functional theory (DFT) treatment for including SOC effects, which was implemented within the \textsc{Crystal} program.\cite{desmarais2021perturbationII,CRYSTAL17PAP,PAPCRYSTAL23} This approach has the potential for over an order of magnitude savings in computation times w.r.t. 2c-SCF, and, moreover, provides a convenient starting point for improvements to treat strongly-correlated, multi-reference, systems, where an ensemble treatment would be necessary.\cite{lieb1983density,gross1988rayleigh,gross1988density,desmarais2021perturbationI,desmarais2021perturbationII,desmaraiskirtmanunpub,helgaker2022lieb} 

Our CPKS treatment is formulated within so-called spin-current DFT (SCDFT), in which the exchange-correlation (xc) functional depends not only on the particle-number density $\rho$ and z-component of the magnetization $m_z$, (as in the usual spin-DFT of von Barth and Hedin),\cite{von1972local} but also on the other Cartesian components of the magnetization $m_x$ and $m_y$, as well as the particle-current $\mathbf{j}$ and spin-current $\mathbf{J}^x$, $\mathbf{J}^y$ and $\mathbf{J}^z$ densities.\cite{w1,w2,vignale1987density,desmarais2019fundamental,desmarais2019spin,desmarais2021spin,desmarais2020adiabatic,bodo2022spin,desmarais2020spin,comaskey2022spin} 

In this paper, we generalize the molecular CPKS treatment to periodic systems. After developing the formalism, our perturbation theory (PT) treatment is implemented in a developmental version of the \textsc{Crystal23} program,\cite{PAPCRYSTAL23} and validated on a family of 2D tungsten dichalcogenide layered compounds, where excellent agreement with reference 2c-SCF calculations is demonstrated, not only for the total energy, but also for band structures and SCDFT density variables.

\section{Formalism}
\label{sec:form}

\subsection{Statement of the Problem}

In the case of periodic systems, the spinors $\vert \psi_{i,\mathbf{k}} \rangle = \vert \psi_{i,\mathbf{k}}^\uparrow \rangle \otimes \vert \uparrow \rangle + \vert \psi_{i,\mathbf{k}}^\downarrow \rangle \otimes \vert \downarrow \rangle$  are 2c crystalline orbitals (COs), with components $\vert \psi_{i,\mathbf{k}}^\sigma \rangle$, expanded in a set of Bloch functions (BFs) $\phi_{\mu,\mathbf{k}}^\sigma$:
\begin{equation}
\label{eqn:co}
\psi_{i,\mathbf{k}}^\sigma \left( \mathbf{r} \right) = \sum_\mu^{N_\mathcal{B}} C^{\sigma}_{\mu,i} \left( \mathbf{k} \right) \phi_{\mu,\mathbf{k}}^\sigma \left( \mathbf{r} \right)
\end{equation}
where $\mathbf{k}$ is a point in the first Brillouin zone (FBZ), $N_\mathcal{B}$ is the number of basis functions in a given cell of the infinite-periodic system and $\sigma=\uparrow, \downarrow$ is a spin index. 

In \textsc{Crystal}, the BFs are conveniently represented as a linear combination of \textit{pure real} atomic orbitals (LCAO), through the inverse-Fourier relation:
\begin{equation}
\label{eqn:Bloch}
\phi_{\mu,\mathbf{k}}^\sigma \left( \mathbf{r} \right) = \frac{1}{\sqrt{\Omega}} \sum_{\mathbf{g}} e^{\imath \mathbf{k} \cdot \mathbf{g}} \ \chi_{\mu,\mathbf{g}}^\sigma \left( \mathbf{r} - \mathbf{A}_\mu \right)
\end{equation}
Here $\Omega$ is the volume of the FBZ, $\mathbf{g}$ is a direct-lattice vector and $\mathbf{A}_\mu$ is the position in cell $\mathbf{g}$ at which the AO $\chi_{\mu,\mathbf{g}}^\sigma$ is centered. In Eq. (\ref{eqn:Bloch}) we have introduced the shorthand notation $\chi_{\mu,\mathbf{g}}^\sigma \left( \mathbf{r} - \mathbf{A}_\mu \right)=\chi_{\mu}^\sigma \left( \mathbf{r} - \mathbf{A}_\mu - \mathbf{g} \right)$. Variation of the orbitals $\psi_{i,\mathbf{k}}^\sigma$, under the constraint of orthonormality:
\begin{subequations}
\begin{equation}
\label{eqn:ortho}
\langle \psi_{i,\mathbf{k}}^\sigma \vert \psi_{j,\mathbf{k}^\prime}^{\sigma^\prime} \rangle = \delta_{i,j} \delta_{\mathbf{k},\mathbf{k}^\prime} \delta_{\sigma,\sigma^\prime} \Rightarrow \mathbf{C}^\dagger \left( \mathbf{k} \right) \mathbf{S} \left( \mathbf{k} \right) \mathbf{C} \left( \mathbf{k} \right) = \mathbf{1}
\end{equation}
leads to the generalized Kohn-Sham (GKS) equation:
\begin{equation}
\label{eqn:GKS}
\mathbf{H} \left( \mathbf{k} \right)  \mathbf{C} \left( \mathbf{k} \right)  = \mathbf{S} \left( \mathbf{k} \right)  \mathbf{C} \left( \mathbf{k} \right)  \mathbf{E} \left( \mathbf{k} \right)
\end{equation}
\end{subequations}
where all matrices have size $2N_\mathcal{B} \times 2N_\mathcal{B}$, $\mathbf{C} \left( \mathbf{k} \right)$ is the matrix of CO coefficients of Eq. (\ref{eqn:co}), $\mathbf{S} \left( \mathbf{k} \right)$ is the BF overlap matrix, $\mathbf{E} \left( \mathbf{k} \right)$ is the matrix of Lagrange multipliers (i.e. for canonical orbitals, corresponding to the diagonal matrix of band-structure energy levels $\epsilon_{i,\mathbf{k}}$) and $\mathbf{H} \left( \mathbf{k} \right)$ is the BF Hamiltonian matrix. Eq. (\ref{eqn:GKS}) can be written more explicitly to highlight the structure in spin space:
\begin{widetext}
\begin{eqnarray}
\label{eqn:GKS_explicit}
\begin{pmatrix}
\mathbf{H}^{\uparrow \uparrow} \left( \mathbf{k} \right) & \mathbf{H}^{\uparrow \downarrow} \left( \mathbf{k} \right)  \\
\mathbf{H}^{\downarrow \uparrow} \left( \mathbf{k} \right) & \mathbf{H}^{\downarrow \downarrow} \left( \mathbf{k} \right)
\end{pmatrix}
\begin{pmatrix}
\mathbf{C}^{\uparrow} \left( \mathbf{k} \right) \\
\mathbf{C}^{\downarrow} \left( \mathbf{k} \right)
\end{pmatrix}
= 
\begin{pmatrix}
\mathbf{S}^{\uparrow \uparrow} \left( \mathbf{k} \right) & \mathbf{0} \\
\mathbf{0} & \mathbf{S}^{\downarrow \downarrow} \left( \mathbf{k} \right)
\end{pmatrix}
\begin{pmatrix}
\mathbf{C}^{\uparrow} \left( \mathbf{k} \right) \\
\mathbf{C}^{\downarrow} \left( \mathbf{k} \right)
\end{pmatrix}
\mathbf{E} \left( \mathbf{k} \right)
\end{eqnarray}
\end{widetext}
In Eq. (\ref{eqn:GKS_explicit}) and elsewhere, matrices with double and single spin indices have size $N_\mathcal{B} \times N_\mathcal{B}$ and $N_\mathcal{B} \times 2N_\mathcal{B}$, respectively. $\mathbf{H}^{\sigma \sigma^\prime} \left( \mathbf{k} \right)$, for instance, has elements:
\begin{equation}
\label{eqn:fock_BF}
H^{\sigma \sigma^\prime}_{\mu \nu} \left( \mathbf{k} \right)= \Omega \langle \phi_{\mu,\mathbf{k}}^\sigma \vert \hat{H} \vert \phi_{\nu,\mathbf{k}}^{\sigma^\prime} \rangle
\end{equation}
and:
\begin{equation}
\label{eqn:fock}
\mathbf{H}^{\sigma \sigma^\prime} \left( \mathbf{k} \right) = \mathbf{h}^{\sigma \sigma^\prime} \left( \mathbf{k} \right) + \mathbf{J}^{\sigma \sigma^\prime} \left( \mathbf{k} \right)- a \mathbf{K}^{\sigma \sigma^\prime} \left( \mathbf{k} \right)+ \mathbf{V}^{\sigma \sigma^\prime} \left( \mathbf{k} \right)\;,
\end{equation}
in which $\mathbf{h}^{\sigma \sigma^\prime} \left( \mathbf{k} \right)$ contains the matrix elements that can be built from mono-electronic integrals:
\begin{equation}
\label{eqn:fock_1e}
\mathbf{h}^{\sigma \sigma^\prime} \left( \mathbf{k} \right) =\delta_{\sigma,\sigma^\prime} \left[ \mathbf{v} \left( \mathbf{k} \right)+ \mathbf{u}_{AR} \left( \mathbf{k} \right)\right]+\mathbf{u}_{SO}^{\sigma \sigma^\prime} \left( \mathbf{k} \right)\;.
\end{equation}
Here, $\mathbf{v}$ consists of the electronic kinetic energy and electron-nuclear interaction terms, $\mathbf{u}_{AR}$ and $\mathbf{u}_{SO}^{\sigma \sigma^\prime}$ are, respectively, the averaged and spin-orbit relativistic effective potential (AREP and SOREP) matrices; and $\mathbf{J}^{\sigma \sigma^\prime}=\delta_{\sigma \sigma^\prime} \boldsymbol{\mathcal{J}}^{\sigma \sigma}$ and $\mathbf{K}^{\sigma \sigma^\prime}$ are the usual Coulomb and exact-exchange terms (with $a$ being the included fraction of the latter). $\mathbf{V}^{\sigma \sigma^\prime}$ is the matrix of DFT correlation and exchange potentials (in either collinear or non-collinear treatments).\cite{desmarais2021perturbationII,desmarais2021spin}

Inserting Eq. (\ref{eqn:Bloch}) into Eq. (\ref{eqn:fock_BF}) (or the equivalent equation with $\hat{H}$ being replaced by any other operator), we are able to relate the BF matrix $\mathbf{H}^{\sigma \sigma^\prime} \left( \mathbf{k} \right)$, for instance, to the AO one $\mathbf{H}^{\sigma \sigma^\prime} \left( \mathbf{g} \right)$ through the inverse-Fourier relation:
\begin{subequations}
\begin{equation}
\label{eqn:fourier}
\mathbf{H}^{\sigma \sigma^\prime} \left( \mathbf{k} \right) = \sum_{\mathbf{g}} e^{\imath \mathbf{k} \cdot \mathbf{g}} \ \mathbf{H}^{\sigma \sigma^\prime} \left( \mathbf{g} \right)
\end{equation}
where AO matrix elements of $\hat{H}$ or any other operator read:
\begin{equation}
\label{eqn:Hg}
\mathbf{H}^{\sigma \sigma^\prime} \left( \mathbf{g} \right) = \langle \chi_{\mu,\mathbf{0}}^\sigma \vert \hat{H} \vert \chi_{\nu,\mathbf{g}}^{\sigma^\prime} \rangle
\end{equation}
\end{subequations}
Following the arguments of Part I,\cite{desmarais2021perturbationI} we recall (assuming real AOs) that the diagonal spin-blocks of $\underline{\mathbf{u}}_{SO} \left( \mathbf{g} \right)$ in Eq. (\ref{eqn:Hg}) are pure imaginary:
\begin{equation}
\label{eqn:uso_imag_g}
\mathcal{R} \left[ \mathbf{u}^{\sigma \sigma}_{SO} \left( \mathbf{g} \right)  \right] =  \mathbf{0} \;,
\end{equation}
whereas the off-diagonal spin-blocks are complex.

In order to develop a computationally convenient PT approach that satisfies Wigner's 2N+1 rule, it turns out to be important to introduce a relation similar to Eq. (\ref{eqn:uso_imag_g}) in the BF basis. This can be achieved by returning to Eq. (\ref{eqn:fourier}) and carrying out the inverse-Fourier transform individually on the real and imaginary parts of $\mathbf{u}^{\sigma \sigma^\prime}_{SO} \left( \mathbf{g} \right)$ and all other matrices entering the Hamiltonian, for instance: 
\begin{subequations}
\begin{equation}
\label{eqn:fourier_r}
\mathbf{H}^{\sigma \sigma^\prime \mathcal{R}} \left( \mathbf{k} \right) = \sum_{\mathbf{g}} e^{\imath \mathbf{k} \cdot \mathbf{g}} \ \Re \left[ \mathbf{H}^{\sigma \sigma^\prime} \left( \mathbf{g} \right) \right]
\end{equation}
and:
\begin{equation}
\label{eqn:fourier_i}
\mathbf{H}^{\sigma \sigma^\prime \mathcal{I}} \left( \mathbf{k} \right) = \imath \sum_{\mathbf{g}} e^{\imath \mathbf{k} \cdot \mathbf{g}} \ \Im \left[ \mathbf{H}^{\sigma \sigma^\prime} \left( \mathbf{g} \right) \right]
\end{equation}
so that:
\begin{equation}
\label{eqn:Hri}
\mathbf{H}^{\sigma \sigma^\prime} \left( \mathbf{k} \right) = \mathbf{H}^{\sigma \sigma^\prime \mathcal{R}} \left( \mathbf{k} \right) + \mathbf{H}^{\sigma \sigma^\prime \mathcal{I}} \left( \mathbf{k} \right)
\end{equation}
\end{subequations}
We note at this point that both $\mathbf{H}^{\sigma \sigma^\prime \mathcal{R}} \left( \mathbf{k} \right)$ and $\mathbf{H}^{\sigma \sigma^\prime \mathcal{I}} \left( \mathbf{k} \right)$ are \textit{complex} quantities. In the case of the SOC operator, using Eq. (\ref{eqn:uso_imag_g}) and proceeding as in Eq. (\ref{eqn:fourier_r}), we obtain:
\begin{equation}
\label{eqn:uso_imag_k}
\mathbf{u}^{\sigma \sigma \mathcal{R}}_{SO} \left( \mathbf{k} \right) =  \mathbf{0} \;.
\end{equation}

\subsection{Order by Order Expressions for the Orthonormality Conditions}

For the expansion of Eq. (\ref{eqn:ortho}) in orders of PT, it proves useful to write the KS coefficients in terms of $N_\mathcal{B} \times N_\mathcal{B}$ blocks with double spin indices:
\footnotesize
\begin{equation}
\label{eqn:CSC_tot}
\begin{pmatrix}
\mathbf{C}^{\uparrow \uparrow} \left( \mathbf{k} \right) & \mathbf{C}^{\downarrow \uparrow} \left( \mathbf{k} \right) \\
\mathbf{C}^{\uparrow \downarrow} \left( \mathbf{k} \right) & \mathbf{C}^{\downarrow \downarrow} \left( \mathbf{k} \right) 
\end{pmatrix}
^\dagger
\begin{pmatrix}
\mathbf{S}^{\uparrow \uparrow} \left( \mathbf{k} \right) & \mathbf{0} \\
\mathbf{0} & \mathbf{S}^{\downarrow \downarrow} \left( \mathbf{k} \right)
\end{pmatrix}
\begin{pmatrix}
\mathbf{C}^{\uparrow \uparrow} \left( \mathbf{k} \right) & \mathbf{C}^{\downarrow \uparrow} \left( \mathbf{k} \right) \\
\mathbf{C}^{\uparrow \downarrow} \left( \mathbf{k} \right) & \mathbf{C}^{\downarrow \downarrow} \left( \mathbf{k} \right) 
\end{pmatrix}
= \mathbf{1}
\end{equation}
\normalsize
As in Part II, we take the scalar-relativistic (SR) unrestricted GKS solution as the zeroth-order problem, which yields:
\begin{equation}
\label{eqn:CSC_0}
\left[ \mathbf{C}^{\sigma \sigma (0)} \left( \mathbf{k} \right) \right]^\dagger \mathbf{S}^{\sigma \sigma} \left( \mathbf{k} \right) \mathbf{C}^{\sigma \sigma (0)} \left( \mathbf{k} \right) = \mathbf{1}
\end{equation}
Then, introducing the $N$th-order orbital rotation matrices $\mathbf{U}^{(N)}$ defined through:
\begin{equation}
\label{eqn:CUC}
\mathbf{C}^{\sigma^\prime \sigma (N)} \left( \mathbf{k} \right) = \mathbf{C}^{\sigma^\prime \sigma^\prime (0)} \left( \mathbf{k} \right) \mathbf{U}^{\sigma^\prime \sigma  (N)} \left( \mathbf{k} \right)
\end{equation}
leads to:
\begin{equation}
\label{eqn:ortho1}
\mathbf{U}^{\sigma^\prime \sigma  (1)} \left( \mathbf{k} \right) = - \left[ \mathbf{U}^{\sigma^\prime \sigma  (1)} \left( \mathbf{k} \right)  \right]^\dagger
\end{equation}
in first order and:
\footnotesize
\begin{equation}
\label{eqn:ortho2}
\mathbf{U}^{\sigma^{\prime} \sigma  (2)} \left( \mathbf{k} \right) + \left[  \mathbf{U}^{\sigma^{\prime} \sigma  (2)} \left( \mathbf{k} \right) \right]^\dagger  = - \sum_{\sigma^{\prime \prime}} \left[ \mathbf{U}^{\sigma^\prime \sigma^{\prime \prime}  (1)} \left( \mathbf{k} \right)  \right]^\dagger \mathbf{U}^{\sigma^{\prime \prime} \sigma  (1)} \left( \mathbf{k} \right)
\end{equation}
\normalsize
in second order.

\subsection{Order by Order Expressions for the Hamiltonian, as well as the Perturbation Equations and their Solution}

For Eq. (\ref{eqn:GKS}) we take the SR unrestricted GKS Hamiltonian as the zeroth-order approximation:
\begin{eqnarray}
\label{eqn:H0}
\mathbf{H}^{\sigma \sigma^\prime (0)} \left( \mathbf{k} \right) &=& \delta_{\sigma \sigma^\prime} \Big[ \mathbf{v} \left( \mathbf{k} \right)+ \mathbf{u}_{AR} \left( \mathbf{k} \right) + \mathbf{J}^{\sigma \sigma (0)} \left( \mathbf{k} \right) \nonumber \\
&-& a \mathbf{K}^{\sigma \sigma (0)} \left( \mathbf{k} \right)+ \mathbf{V}^{\sigma \sigma (0)} \left( \mathbf{k} \right) \Big]
\end{eqnarray}
In first order:
\begin{equation}
\label{eqn:H1}
\mathbf{H}^{\sigma \sigma^\prime (1)} \left( \mathbf{k} \right) = \mathbf{u}_{SO}^{\sigma \sigma^\prime} \left( \mathbf{k} \right) + \mathbf{J}^{\sigma \sigma^\prime (1)} \left( \mathbf{k} \right)- a \mathbf{K}^{\sigma \sigma^\prime (1)} \left( \mathbf{k} \right)+ \mathbf{V}^{\sigma \sigma^\prime (1)} \left( \mathbf{k} \right)
\end{equation}
and for all orders $N$ greater than one:
\begin{equation}
\label{eqn:HN}
\mathbf{H}^{\sigma \sigma^\prime (N)} \left( \mathbf{k} \right) = \mathbf{J}^{\sigma \sigma^\prime (N)} \left( \mathbf{k} \right)- a \mathbf{K}^{\sigma \sigma^\prime (N)} \left( \mathbf{k} \right)+ \mathbf{V}^{\sigma \sigma^\prime (N)} \left( \mathbf{k} \right)
\end{equation}
Then, following Part II, expansion of Eq. (\ref{eqn:GKS}) in orders of PT leads, in the first order, to:
\begin{equation}
\label{eqn:pert_1}
\mathbf{G}^{(1)} \left( \mathbf{k} \right)  + \mathbf{E}^{(0)} \left( \mathbf{k} \right) \ \mathbf{U}^{(1)} \left( \mathbf{k} \right) = \mathbf{U}^{(1)} \left( \mathbf{k} \right) \ \mathbf{E}^{(0)} \left( \mathbf{k} \right) + \mathbf{E}^{(1)}\left( \mathbf{k} \right) \;,
\end{equation}
and, in second order, to:
\begin{widetext}
\begin{eqnarray}
\label{eqn:pert_2}
\mathbf{G}^{(2)} \left( \mathbf{k} \right) + \mathbf{E}^{(0)} \left( \mathbf{k} \right) \ \mathbf{U}^{(2)} \left( \mathbf{k} \right) + \mathbf{G}^{(1)} \left( \mathbf{k} \right) \ \mathbf{U}^{(1)} \left( \mathbf{k} \right) = 
\mathbf{U}^{(2)} \left( \mathbf{k} \right) \ \mathbf{E}^{(0)} \left( \mathbf{k} \right) + \mathbf{U}^{(1)} \left( \mathbf{k} \right) \ \mathbf{E}^{(1)} \left( \mathbf{k} \right) + \mathbf{E}^{(2)} \left( \mathbf{k} \right) \;,
\end{eqnarray}
\end{widetext}
where we have made use of the $N$th order Hamiltonian matrix in the CO basis:
\begin{equation}
\label{eqn:GN}
\mathbf{G}^{(N)} \left( \mathbf{k} \right)  = \left[ \mathbf{C}^{(0)} \left( \mathbf{k} \right) \right]^\dagger \mathbf{H}^{(N)} \left( \mathbf{k} \right) \mathbf{C}^{(0)} \left( \mathbf{k} \right)
\end{equation}

For further development, and following Eq. (\ref{eqn:Hri}), it is useful to define:
\begin{subequations}
\begin{equation}
\label{eqn:Gri}
\mathbf{G}^{(N)} \left( \mathbf{k} \right)  = \mathbf{G}^{\mathcal{R} (N)} \left( \mathbf{k} \right) + \mathbf{G}^{\mathcal{I} (N)} \left( \mathbf{k} \right) 
\end{equation}
as well as:
\begin{equation}
\label{eqn:Uri}
\mathbf{U}^{(N)} \left( \mathbf{k} \right)  = \mathbf{U}^{\mathcal{R} (N)} \left( \mathbf{k} \right) + \mathbf{U}^{\mathcal{I} (N)} \left( \mathbf{k} \right) 
\end{equation}
where:
\begin{equation}
\label{eqn:Gc}
\mathbf{G}^{\mathcal{C} (N)} \left( \mathbf{k} \right) = \left[ \mathbf{C}^{(0)} \left( \mathbf{k} \right) \right]^\dagger \mathbf{H}^{\mathcal{C} (N)} \left( \mathbf{k} \right) \mathbf{C}^{(0)} \left( \mathbf{k} \right)
\end{equation}
and $\mathcal{C} = \mathcal{R} \text{ or } \mathcal{I}$.
\end{subequations}
Recalling from Eq. (\ref{eqn:uso_imag_k}) that $\mathbf{u}^{\sigma \sigma \mathcal{R}}_{SO} \left( \mathbf{k} \right)$ is zero, the Coulomb and exchange response to a vanishing perturbation is also vanishing. Thus, inserting Eqs. (\ref{eqn:uso_imag_k}) and (\ref{eqn:H1}) in Eq. (\ref{eqn:Gc}), we obtain, for diagonal spin blocks:
\begin{equation}
\label{eqn:imag_Gr}
\mathbf{G}^{\sigma \sigma  \mathcal{R} (1)} \left( \mathbf{k} \right) = \mathbf{0}
\end{equation}
As will be discussed below in Section \ref{sec:k_mk}, partitioning of matrices in $\mathcal{R}$ and $\mathcal{I}$ contributions, as in Eq. (\ref{eqn:Gri}) and  (\ref{eqn:Uri}) allows us to relate elements upon inversion in reciprocal space.

As far as the GKS perturbed wave functions and band structures are concerned, again following Part II, we solve Eqs. (\ref{eqn:ortho1}), (\ref{eqn:ortho2}),  (\ref{eqn:pert_1}) and (\ref{eqn:pert_2}). In doing so, we take advantage of the fact that the off-diagonal occupied-virtual blocks of the Lagrange-multiplier matrix vanish at all orders $N$:
\begin{equation}
\label{eqn:eps_VO}
\mathbf{E}_{VO}^{\sigma \sigma^\prime (N)} \left( \mathbf{k} \right) = \mathbf{E}_{OV}^{\sigma^\prime \sigma (N)} \left( \mathbf{k} \right) = \mathbf{0}
\end{equation}
In first order, the solution of Eqs. (\ref{eqn:ortho1}) and (\ref{eqn:pert_1}), together with Eq. (\ref{eqn:imag_Gr}) provides, for the virtual-occupied (VO) block:
\begin{equation}
\label{eqn:U1C_VO}
\mathbf{U}^{\sigma \sigma^\prime \mathcal{C} (1)}_{VO} \left( \mathbf{k} \right) = \frac{\mathbf{G}^{\sigma \sigma^\prime  \mathcal{C} (1)}_{VO} \left( \mathbf{k} \right)  }{ \epsilon_O^{\sigma^\prime(0)} \left( \mathbf{k} \right)  - \epsilon_V^{\sigma(0)} \left( \mathbf{k} \right)  } = - \left[ \mathbf{U}^{\sigma^\prime \sigma \mathcal{C} (1)}_{OV} \left( \mathbf{k} \right)  \right]^\ast
\end{equation}
while for the occupied-occupied (OO$^\prime$) and virtual-virtual (VV$^\prime$) blocks:
\begin{equation}
\label{eqn:U1C_OO_VV}
\mathbf{U}^{\sigma \sigma^\prime \mathcal{C} (1)}_{OO^\prime} \left( \mathbf{k} \right) = \mathbf{U}^{\sigma \sigma^\prime \mathcal{C} (1)}_{VV^\prime} \left( \mathbf{k} \right) = \mathbf{0}
\end{equation}
in the non-canonical treatment. It follows that, for the occupied-occupied block:
\begin{equation}
\label{eqn:eps_OO}
\mathbf{E}_{OO^\prime}^{\sigma \sigma^\prime (1)} \left( \mathbf{k} \right) = \mathbf{G}^{\sigma \sigma^\prime \mathcal{R} (1)}_{OO^\prime} \left( \mathbf{k} \right) + \mathbf{G}^{\sigma \sigma^\prime \mathcal{I} (1)}_{OO^\prime} \left( \mathbf{k} \right)
\end{equation}
with an exactly analogous expression also for the virtual-virtual block $\mathbf{E}_{VV^\prime}^{\sigma \sigma^\prime (1)} \left( \mathbf{k} \right)$. A single diagonalization of, for instance, $\mathbf{E}_{OO^\prime}^{\sigma \sigma^\prime (0)} \left( \mathbf{k} \right)+\mathbf{E}_{OO^\prime}^{\sigma \sigma^\prime (1)} \left( \mathbf{k} \right)$ at the end of the calculation (i.e. after a coupled solution of Eq. \ref{eqn:pert_1}) yields the occupied band structure correct through first-order in PT.

At second order, we proceed similarly and write $\mathbf{U}^{(2)}$ as in Eq. (\ref{eqn:Uri}) in terms of contributions $\mathbf{U}^{\mathcal{R}(2)}$, and $\mathbf{U}^{\mathcal{I}(2)}$ that end up having different behaviour upon inversion in reciprocal space - cf. Eq. (\ref{eqn:UN_k_mk}). Then, the simultaneous non-canonical solution of Eqs. (\ref{eqn:pert_2}) and (\ref{eqn:ortho2}) yields for the occupied-occupied blocks:
\begin{subequations}
\begin{eqnarray}
\label{eqn:UooR2}
\mathbf{U}^{\sigma \sigma^\prime \mathcal{R} (2)}_{OO^\prime} \left( \mathbf{k} \right)= \left[ \mathbf{U}^{\sigma^\prime \sigma \mathcal{R} (2)}_{O^\prime O} \left( \mathbf{k} \right) \right]^\ast = \nonumber \\
 -\frac{1}{2} \sum_{\mathcal{C}} \sum_{\sigma^{\prime \prime}}  \sum_{V} \left[ \mathbf{U}^{\sigma^{\prime \prime}\sigma \mathcal{C} (1)}_{VO} \left( \mathbf{k} \right) \right]^\ast \mathbf{U}^{\sigma^{\prime \prime} \sigma^\prime \mathcal{C} (1)}_{V O^\prime} \left( \mathbf{k} \right)
\end{eqnarray}
and:
\begin{eqnarray}
\label{eqn:UooI2}
\mathbf{U}^{\sigma \sigma^\prime \mathcal{I} (2)}_{OO^\prime} \left( \mathbf{k} \right)= \left[ \mathbf{U}^{\sigma^\prime \sigma \mathcal{I} (2)}_{O^\prime O} \left( \mathbf{k} \right) \right]^\ast= \nonumber \\
 -\frac{1}{2} \sum_{\mathcal{C}} \sum_{\mathcal{C}^\prime \ne \mathcal{C}} \sum_{\sigma^{\prime \prime}}  \sum_{V} \left[ \mathbf{U}^{\sigma^{\prime \prime}\sigma \mathcal{C} (1)}_{VO} \left( \mathbf{k} \right) \right]^\ast \mathbf{U}^{\sigma^{\prime \prime} \sigma^\prime \mathcal{C}^\prime (1)}_{V O^\prime} \left( \mathbf{k} \right) \nonumber \\
\end{eqnarray}
\end{subequations}
Exactly analogous expressions can also be determined for $\mathbf{U}^{\sigma \sigma^\prime \mathcal{R} (2)}_{VV^\prime} \left( \mathbf{k} \right)$ and $\mathbf{U}^{\sigma \sigma^\prime \mathcal{I} (2)}_{VV^\prime} \left( \mathbf{k} \right)$. To obtain Eqs. (\ref{eqn:UooR2}) and (\ref{eqn:UooI2}), we insert Eqs. (\ref{eqn:U1C_VO}) and (\ref{eqn:U1C_OO_VV}) into Eq. (\ref{eqn:ortho2}). The expressions for the VO blocks, on the other hand are provided in Appendix \ref{app:U2vo}. In Eqs. (\ref{eqn:UooR2}), (\ref{eqn:UooI2}) and elsewhere the products inside sums over $O$ or $V$ are performed element-wise.

Eqs. (\ref{eqn:UooR2}), (\ref{eqn:UooI2}),  (\ref{eqn:UvoR2}) and (\ref{eqn:UvoI2}) are consistent with a set of non-canonical second-order Lagrange multipliers that are provided in Appendix \ref{app:E2}. A single diagonalization of, for instance, $\mathbf{E}_{OO^\prime}^{\sigma \sigma^\prime (0)} \left( \mathbf{k} \right)+\mathbf{E}_{OO^\prime}^{\sigma \sigma^\prime (1)} \left( \mathbf{k} \right) + \mathbf{E}_{OO^\prime}^{\sigma \sigma^\prime (2)} \left( \mathbf{k} \right)$ at the end of the calculation (i.e. after a coupled solution of Eqs. \ref{eqn:pert_1} and \ref{eqn:pert_2}) provides occupied band structure energy levels that are correct through second-order in PT.

As will be shown in Section \ref{sec:res}, it turns out that an excellent approximation to the full solution of Eq. (\ref{eqn:GKS}) is given by the coupled solution of Eq. (\ref{eqn:pert_1}), followed by an uncoupled solution of Eq. (\ref{eqn:pert_2}). Such a procedure, which we call the PT2$^\prime$  approximation is mathematically achieved by simply setting:
\begin{equation}
\label{eqn:pt2_primo}
\text{PT2$^\prime$ ::} \quad \mathbf{G}^{\mathcal{R} (2)} \left( \mathbf{k} \right) , \mathbf{G}^{\mathcal{I} (2)}\left( \mathbf{k} \right) \Rightarrow \mathbf{0}
\end{equation}
in Eqs. (\ref{eqn:UvoR2}), (\ref{eqn:UvoI2}), as well as Eqs. (\ref{eqn:eps2_oor}) and (\ref{eqn:eps2_ooi}).

\subsection{Some $\mathbf{k}$ to $-\mathbf{k}$ Relations}

\label{sec:k_mk}
Given that the zeroth-order direct-space SR Hamiltonian, being the inverse-Fourier transform of Eq. (\ref{eqn:H0}), is pure-real, we obtain:
\begin{equation}
\label{eqn:H0_k_mk}
\mathbf{H}^{\sigma \sigma^\prime (0)} \left( \mathbf{k} \right) = \sum_{\mathbf{g}} e^{\imath \mathbf{k} \cdot \mathbf{g}} \ \mathbf{H}^{\sigma \sigma^\prime (0)} \left( \mathbf{g} \right) = \left[ \mathbf{H}^{\sigma \sigma^\prime (0)} \left( - \mathbf{k} \right) \right]^\ast
\end{equation}
This leads to the following well-known result for the CO coefficient and Lagrange multiplier matrices at $\mathbf{k}$ and $-\mathbf{k}$:\cite{desmarais2020spin}
\begin{subequations}
\begin{equation}
\label{eqn:c_k_mk}
\mathbf{C}^{(0)} \left( \mathbf{k} \right) = \left[ \mathbf{C}^{(0)} \left( -\mathbf{k} \right) \right]^\ast
\end{equation}
and:
\begin{equation}
\label{eqn:e_k_mk}
\mathbf{E}^{(0)} \left( \mathbf{k} \right) = \mathbf{E}^{(0)} \left( -\mathbf{k} \right)
\end{equation}
\end{subequations}
Proceeding as in Eq. (\ref{eqn:H0_k_mk}) at order $N > 0$, we obtain:
\small
\begin{subequations}
\label{eqn:HN_k_mk}
\begin{eqnarray}
\mathbf{H}^{\sigma \sigma^\prime \mathcal{R}(N)} \left( \mathbf{k} \right) &=& \sum_{\mathbf{g}} e^{\imath \mathbf{k} \cdot \mathbf{g}} \ \Re \left[ \mathbf{H}^{\sigma \sigma^\prime} \left( \mathbf{g} \right) \right] =  \left[ \mathbf{H}^{\sigma \sigma^\prime \mathcal{R}(N)} \left( -\mathbf{k} \right)  \right]^\ast \\
\mathbf{H}^{\sigma \sigma^\prime \mathcal{I}(N)} \left( \mathbf{k} \right) &=& \imath \sum_{\mathbf{g}} e^{\imath \mathbf{k} \cdot \mathbf{g}} \ \Im \left[ \mathbf{H}^{\sigma \sigma^\prime} \left( \mathbf{g} \right) \right] =  - \left[ \mathbf{H}^{\sigma \sigma^\prime \mathcal{I}(N)} \left( -\mathbf{k} \right)  \right]^\ast
\end{eqnarray}
\end{subequations}
\normalsize
Then, inserting Eqs. (\ref{eqn:HN_k_mk}) and (\ref{eqn:c_k_mk}) into Eq. (\ref{eqn:Gc}) gives:
\begin{subequations}
\label{eqn:GN_k_mk}
\begin{eqnarray}
\mathbf{G}^{\sigma \sigma^\prime \mathcal{R}(N)} \left( \mathbf{k} \right) &=& \left[ \mathbf{G}^{\sigma \sigma^\prime \mathcal{R}(N)} \left( -\mathbf{k} \right)  \right]^\ast \\
\mathbf{G}^{\sigma \sigma^\prime \mathcal{I}(N)} \left( \mathbf{k} \right) &=& - \left[ \mathbf{G}^{\sigma \sigma^\prime \mathcal{I}(N)} \left( -\mathbf{k} \right)  \right]^\ast
\end{eqnarray}
\end{subequations}
and then, inserting Eqs. (\ref{eqn:e_k_mk}) and (\ref{eqn:GN_k_mk}) into Eqs. (\ref{eqn:U1C_VO})-(\ref{eqn:U1C_OO_VV}) as well as Eqs. (\ref{eqn:UooR2}), (\ref{eqn:UooI2}),  (\ref{eqn:UvoR2}) and (\ref{eqn:UvoI2}) gives:
\begin{subequations}
\label{eqn:UN_k_mk}
\begin{eqnarray}
\mathbf{U}^{\sigma \sigma^\prime \mathcal{R}(N)} \left( \mathbf{k} \right) &=& \left[ \mathbf{U}^{\sigma \sigma^\prime \mathcal{R}(N)} \left( -\mathbf{k} \right)  \right]^\ast \\
\mathbf{U}^{\sigma \sigma^\prime \mathcal{I}(N)} \left( \mathbf{k} \right) &=& - \left[ \mathbf{U}^{\sigma \sigma^\prime \mathcal{I}(N)} \left( -\mathbf{k} \right)  \right]^\ast
\end{eqnarray}
\end{subequations}

\subsection{Order by Order Expressions for the Direct-Space Density Matrix}

The GKS density matrix, obtained as a solution of Eq. (\ref{eqn:GKS}) reads:
\begin{equation}
\label{eqn:P}
\mathbf{P}^{\sigma \sigma^\prime} \left( \mathbf{k} \right) = \sum_{\sigma^{\prime \prime}} \mathbf{C}^{\sigma \sigma^{\prime \prime}} \left( \mathbf{k} \right)  \mathbf{f}_{\sigma^{\prime \prime}} \left( \mathbf{k} \right) \left[ \mathbf{C}^{\sigma^{\prime \prime} \sigma^\prime} \left( \mathbf{k} \right) \right]^\dagger
\end{equation}
where $\mathbf{f}_{\sigma}\left( \mathbf{k} \right)$ is the diagonal matrix of band occupations (i.e. for insulators the matrix elements are $1$ for occupied bands and $0$ for virtual bands, or, in general fractional values for partially occupied bands in metals). Expanding the CO coefficients in orders of perturbation theory, using Eqs. (\ref{eqn:CUC}), we obtain, for instance, in zeroth order:
\begin{subequations}
\begin{equation}
\label{eqn:P0}
\mathbf{P}^{\sigma \sigma^\prime (0)} \left( \mathbf{k} \right) = \delta_{\sigma \sigma^\prime} \mathbf{C}^{\sigma \sigma (0)} \left( \mathbf{k} \right) \mathbf{f}_{\sigma} (\mathbf{k}) \left[ \mathbf{C}^{\sigma \sigma (0)} \left( \mathbf{k} \right) \right]^\dagger
\end{equation}
in first order:
\begin{eqnarray}
\label{eqn:P1}
\mathbf{P}^{\sigma \sigma^\prime(1)} ( \mathbf{k} ) &=& \mathbf{C}^{\sigma \sigma(0)} (\mathbf{k}) \mathbf{f}_\sigma (\mathbf{k}) \left[ \mathbf{U}^{\sigma \sigma^\prime(1)} (\mathbf{k}) \right]^\dagger \left[ \mathbf{C}^{\sigma^\prime \sigma^\prime (0)} (\mathbf{k}) \right]^\dagger \nonumber \\
&+& \mathbf{C}^{\sigma \sigma(0)} (\mathbf{k}) \mathbf{U}^{\sigma \sigma^\prime(1)} (\mathbf{k}) \mathbf{f}_{\sigma^\prime} (\mathbf{k}) \left[ \mathbf{C}^{\sigma^\prime \sigma^\prime (0)} (\mathbf{k}) \right]^\dagger
\end{eqnarray}
and, in second order, an expression that is provided in Appendix \ref{app:pg}.
\end{subequations}
The matrices in direct space are obtained from an inverse-Fourier transform:
\begin{widetext}
\begin{eqnarray}
\label{eqn:p1_g1}
\mathbf{P}^{\sigma \sigma^\prime(N)} ( \mathbf{g} ) = \frac{1}{\Omega} \int d \mathbf{k} \ e^{\imath \mathbf{k} \cdot \mathbf{g}} \ \mathbf{P}^{\sigma \sigma^\prime(N)} ( \mathbf{k} ) = \frac{1}{\Omega} \int^\prime d \mathbf{k} 
\Big\{  e^{\imath \mathbf{k} \cdot \mathbf{g}} \mathbf{P}^{\sigma \sigma^\prime(N)} ( \mathbf{k} ) + e^{-\imath \mathbf{k} \cdot \mathbf{g}} \mathbf{P}^{\sigma \sigma^\prime(N)} ( -\mathbf{k} ) \Big\}
\end{eqnarray}
\end{widetext}
where the prime over the integral indicates that integration is limited to the portion of the FBZ with positive coordinates. Inserting Eqs. (\ref{eqn:U1C_VO}),  (\ref{eqn:U1C_OO_VV}) and (\ref{eqn:p1_g1}) into Eq. (\ref{eqn:P1}) and making use of Eq. (\ref{eqn:c_k_mk}) and (\ref{eqn:UN_k_mk}) to relate elements at $\mathbf{k}$ and $-\mathbf{k}$, gives, in first order, for the case of diagonal spin-blocks of $\mathbf{P}^{(1)}$ (see Appendix \ref{app:pg} for details):
\begin{equation}
\label{eqn:rp1_0}
\Re \left[ \mathbf{P}^{\sigma \sigma (1)} ( \mathbf{g} ) \right] = \mathbf{0} \\
\end{equation}
and
\begin{widetext}
\begin{eqnarray}
\label{eqn:ip1}
\Im \left[ \mathbf{P}^{\sigma \sigma (1)} ( \mathbf{g} ) \right] =  \frac{2}{\Omega} \Im \int^\prime \!\! d \mathbf{k} 
\Big\{ e^{\imath \mathbf{k} \cdot \mathbf{g}} \mathbf{C}^{\sigma \sigma(0)}_O (\mathbf{k}) \mathbf{f}_{\sigma O} (\mathbf{k}) \left[ \mathbf{U}^{\sigma \sigma \mathcal{I} (1)}_{OV} (\mathbf{k}) \right]^\dagger \!\!\left[ \mathbf{C}^{\sigma \sigma (0)}_V (\mathbf{k}) \right]^\dagger \!\!
+ e^{\imath \mathbf{k} \cdot \mathbf{g}} \mathbf{C}^{\sigma \sigma(0)}_V (\mathbf{k}) \mathbf{U}^{\sigma \sigma \mathcal{I} (1)}_{VO} (\mathbf{k}) \mathbf{f}_{\sigma O} (\mathbf{k}) \left[ \mathbf{C}^{\sigma \sigma (0)}_O (\mathbf{k}) \right]^\dagger \!\!\Big\} 
\end{eqnarray}
\end{widetext}
The expressions for the full $\mathbf{P}^{(1)} ( \mathbf{g} ) $ and $\mathbf{P}^{(2)} ( \mathbf{g} )$ are provided in Appendix \ref{app:pg}. To obtain Eqs. (\ref{eqn:rp1_0}) and (\ref{eqn:ip1}), we insert Eqs. (\ref{eqn:imag_Gr}) and (\ref{eqn:U1C_VO}) into Eq. (\ref{eqn:p1_g}). As was the case in Part II of this series,\cite{desmarais2021perturbationII} the vanishing $\Re \left[ \mathbf{P}^{\sigma \sigma (1)} ( \mathbf{g} ) \right]$ leads to a vanishing first-order electron-density response, and thus a vanishing Coulomb and exchange-correlation response (for functionals not depending on the particle- and spin-current densities). This leads to an uncoupled-perturbed procedure for pure LDA and GGA functionals, and a coupled-perturbed procedure for hybrid functionals, including a fraction of exact Fock exchange.

\subsection{Order by Order Expressions for the Total Energy}

Having set up all pieces of the PT, we are ready to derive computationally convenient expressions for the total energy that are consistent with Wigner's 2N+1 rule. The full derivation, which also resembles that given in Part II of this series,\cite{desmarais2021perturbationII} is provided in the electronic supporting information (ESI).\cite{ESI_PT3} Here we report the final expressions. In first order, the energy corrections due to SOC are vanishing:
\begin{equation}
\label{eqn:e1}
E^{(1)} = 0
\end{equation}
For second and third order, it proves convenient to introduce the matrix of SOC integrals in the CO basis:
\begin{equation}
\boldsymbol{\Xi}^{\sigma \sigma^\prime \mathcal{C}} ( \mathbf{k} ) = \left[ \mathbf{C}^{\sigma \sigma(0)} (\mathbf{k}) \right]^\dagger \mathbf{u}^{\sigma \sigma^\prime \mathcal{C}}_{SO} \left( \mathbf{k} \right) \mathbf{C}^{\sigma^\prime \sigma^\prime  (0)} (\mathbf{k})
\end{equation}
in terms of which we obtain:
\begin{equation}
\label{eqn:e2}
E^{(2)} = \frac{2}{\Omega} \int^\prime d \mathbf{k}  \sum_{\mathcal{C}} \sum_{\sigma \sigma^\prime} \Re \text{Tr} \{ \mathbf{f}_{\sigma O} (\mathbf{k}) \boldsymbol{\Xi}^{\sigma \sigma^\prime \mathcal{C}}_{OV} ( \mathbf{k} ) \mathbf{U}^{\sigma^\prime \sigma \mathcal{C} (1)}_{VO} (\mathbf{k}) \}
\end{equation}
and
\begin{widetext}
\begin{eqnarray}
\label{eqn:e3_3}
E^{(3)} &=& \frac{2}{\Omega} \int^\prime d \mathbf{k}  \sum_{\mathcal{C},\mathcal{C}^\prime} \SUMP{\mathcal{C}^{\prime \prime}}{}  \sum_{\sigma \sigma^\prime \sigma^{\prime \prime} } \Re \text{Tr} \Big\{ \mathbf{G}^{\sigma \sigma^\prime \mathcal{C}  (1)}_{VV^\prime} ( \mathbf{k} ) \mathbf{U}^{\sigma^\prime \sigma^{\prime \prime} \mathcal{C}^\prime  (1)}_{V^\prime O} (\mathbf{k}) \mathbf{f}_{\sigma^{\prime \prime} O} (\mathbf{k}) \left[ \mathbf{U}^{\sigma^{\prime \prime} \sigma  \mathcal{C}^{\prime \prime}  (1)}_{OV} (\mathbf{k}) \right]^\dagger \nonumber \\
&-& \frac{1}{2} \left[ \mathbf{f}_{\sigma O} (\mathbf{k}) \mathbf{G}^{\sigma \sigma^\prime \mathcal{C}(1)}_{OO^\prime } (\mathbf{k})  +  \mathbf{G}^{\sigma \sigma^\prime \mathcal{C}(1)}_{OO^\prime } (\mathbf{k}) \mathbf{f}_{\sigma^\prime O^\prime } (\mathbf{k}) \right] \left[ \mathbf{U}^{\sigma^\prime \sigma^{\prime \prime} \mathcal{C}^\prime  (1)}_{O^\prime V} (\mathbf{k}) \right]^\dagger \mathbf{U}^{\sigma^{\prime \prime} \sigma \mathcal{C}^{\prime \prime}  (1)}_{VO } (\mathbf{k}) \Big\}
\end{eqnarray}
\end{widetext}
where the prime over the sum on $\mathcal{C}^{\prime \prime}$ means that $\mathcal{C}^{\prime \prime}=\mathcal{R}$ if $\mathcal{C} = \mathcal{C}^\prime$ and, otherwise $\mathcal{C}^{\prime \prime}=\mathcal{I}$. Eqs. (\ref{eqn:e1}) to (\ref{eqn:e3_3}) show that the energy may be determined up to third order, using only the first-order perturbed wavefunction.

\begin{figure}[t!]
\centering
\includegraphics[width=8.6cm]{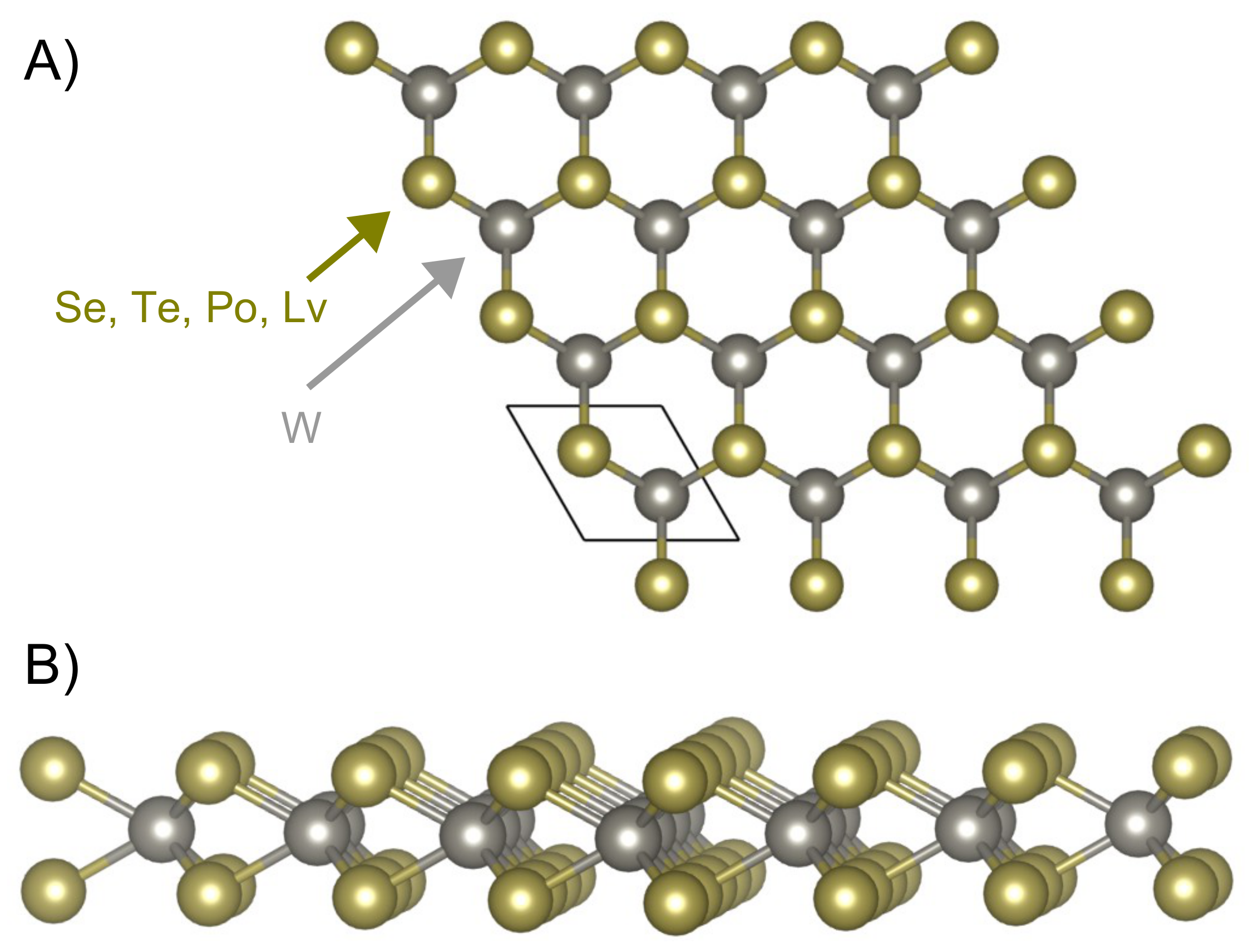}
\caption{Graphical representation of the atomic structure of the series of tungsten dichalcogenide 2D layered systems considered in this study to test the proposed perturbation theory treatment of SOC: WX$_2$ with X = Se, Te, Po, Lv.}
\label{fig:sys}
\end{figure}

\section{Computational Details}
\label{sec:comp}

Calculations are performed on tungsten dichalcogenide 2D layers, shown in Figure \ref{fig:sys}, which have been chosen based on their previously reported ``giant spin-orbit induced spin-splitting''.\cite{zhu2011giant} All calculations are performed with the PBE0 global hybrid functional.\cite{pbe0_art} We did not test other functionals because the previous study on molecular systems has already shown that the convergence behaviour of the perturbation series is very similar with five different functionals.\cite{desmarais2021perturbationII} The present calculations employed the small-core \texttt{STUTSC} potential (for W and Lv) and both the large- and small-core \texttt{STUTSC}, as well as \texttt{STUTLC} potentials for Se, Te, Po. For W, the valence basis set was of the form (6$s$6$p$4$d$2$f$)/[5$s$3$p$4$d$2$f$], being modified starting from the ecp-60-dhf-SVP set available from the \textsc{Turbomole} package.\cite{TURBOMOLE} For the \texttt{STUTLC} calculations, the valence basis sets for Se, Te and Po of the form (5$s$5$p$2$d$)/[3$s$3$p$2$d$] were modified from the ones originally presented in Ref. \onlinecite{stoll2002relativistic}. For the \texttt{STUTSC} calculations, the valence basis sets for Se, Te and Po are of the form (18$s$15$p$8$d$)/[4$s$3$p$2$d$] or (18$s$13$p$7$d$)/[4$s$3$p$2$d$]. The valence basis set for Lv is an uncontracted one of the form (10$s$8$p$7$d$1$f$)/[10$s$8$p$7$d$1$f$]. The full input decks are available in \textsc{Crystal} format in the ESI.\cite{ESI_PT3} Reciprocal space was sampled in a $24 \times 24$ Monkhorst-Pack net, with Fermi smearing of 0.001 $E_h$. A tolerance of 10$^{-8}$ $E_h$ on the total energy was used as a convergence criterion for the SCF procedure. The five \texttt{TOLINTEG} parameters that control truncation of the Coulomb and exact-exchange infinite series were set to 8 8 8 8 30. The exchange-correlation functional and potential (in their collinear spin-DFT formulation) were sampled on a direct-space pruned grid over the unit-cell volume with Lebedev angular and Gauss-Legendre radial quadratures, employing 99 radial and 1454 angular points (keyword \texttt{XXLGRID}). The geometries of the layers were initially obtained by cleaving three-atom thick slabs along the (001) surface of the bulk P6$_3$/mmc crystal structures.\cite{schutte1987crystal} Then, both the atomic fractional coordinates and lattice parameters of the layers were fully optimized with analytical gradients of the total energy for systems periodic in two dimensions, and a quasi-Newton scheme, using, respectively, the PBE and PBE0 functionals at the scalar-relativistic 1c-SCF level.\cite{Doll1,Doll2,DOLL_CELLGRAD2,CivaGRAD} Finally, single-point 2c-SCF and CPKS calculations, including SOC, were performed on the previously optimized scalar-relativistic geometries.

\section{Results and Discussion}
\label{sec:res}

In the following, we validate the present pure-state periodic CPKS approach for SOC against reference periodic 2c-SCF calculations.\cite{desmarais2020spin} Comparisons are reported on i) total energies, ii) electronic band structures, and iii) spatial distribution of density variables of spin-current DFT.

\small
\begin{table}[b]
\caption{Percentage contribution of SOC to the energy with large-core (LC) and small-core (SC) RECPs. Relative percentage differences w.r.t. 2c-SCF values are also reported.}
\label{tab:energy_analysis}
\vspace{5pt}
\begin{tabular}{llcccccc}
\hline
\hline
&&\\
& & $\%^\text{PT2}$ & $\%^\text{PT3}$ & $\%^\text{2c}$ & & $\vert \Delta\%_2 \vert$ & $\vert \Delta\%_3 \vert$\\
\hline
&&\\
\multirow{3}{*}{LC} & $\text{WSe}_2$ & $2.26\text{E}^{-2}$ & $2.20\text{E}^{-2}$ & $2.21\text{E}^{-2}$ & & $2.17$ & $0.50$ \\
& $\text{WTe}_2$ & $2.97\text{E}^{-2}$ & $2.92\text{E}^{-2}$ & $2.93\text{E}^{-2}$ & & $1.23$ & $0.51$ \\
& $\text{WPo}_2$ & $7.97\text{E}^{-2}$ & $8.12\text{E}^{-2}$ & $8.13\text{E}^{-2}$ & & $1.93$ & $0.13$ \\
&&\\
\multirow{4}{*}{SC} & $\text{WSe}_2$ & $2.69\text{E}^{-3}$ & $2.62\text{E}^{-3}$ & $2.63\text{E}^{-3}$ & & $2.14$ & $0.46$ \\
&$\text{WTe}_2$ & $6.18\text{E}^{-3}$ & $6.01\text{E}^{-3}$ & $6.03\text{E}^{-3}$ & & $2.43$ & $0.34$ \\
& $\text{WPo}_2$ & $3.37\text{E}^{-2}$ & $3.20\text{E}^{-2}$ & $3.20\text{E}^{-2}$ & & $5.16$ & $0.05$ \\
& $\text{WLv}_2$ & $6.57\text{E}^{-1}$ & $7.22\text{E}^{-1}$ & $7.10\text{E}^{-1}$ & & $7.43$ & $1.77$ \\
&&\\
\hline 
\hline
\end{tabular}
\end{table}
\normalsize

\subsection{Total Energies}

Table \ref{tab:energy_analysis} reports percentage contributions of SOC to total energies through second and third order in PT ($\%^\text{PT2}$ and $\%^\text{PT3}$), as well as with the 2c-SCF reference method ($\%^\text{2c}$). Relative differences w.r.t. 2c-SCF are also provided ($\vert \Delta\%_2 \vert$ and $\vert \Delta\%_3 \vert$ through second and third order, respectively). It is found that third order contributions to the total energy consistently improve the agreement w.r.t 2c-SCF, as $\vert \Delta\%_3 \vert$ is always smaller than $\vert \Delta\%_2 \vert$. For $\text{WSe}_2$ the error in predicting the SOC contribution to the energy through second order in PT is $\vert \Delta\%_2 \vert=2.17 \%$ with the LC potential or $2.14 \%$ with the SC potential. These figures are improved to $\vert \Delta\%_3 \vert=0.5 \%$ and $\vert \Delta\%_3 \vert=0.46 \%$  through third-order. In fact, an error larger than $1\%$ through third order, is only found on $\text{WLv}_2$, with $\vert \Delta\%_3 \vert=1.77\%$, being a large improvement over the second-order value of $\vert \Delta\%_2 \vert=7.43\%$.

\subsection{Comparison of PT2$^\prime$ vs. PT2 Approximations}

Before discussing the comparisons of the perturbation theories for one-electron properties on periodic systems, we first validate the PT2$^\prime$ approximation for GKS eigenvalues and density variables of SCDFT of Eq. (\ref{eqn:pt2_primo}) on the well-studied molecular systems of Part II.\cite{desmarais2021perturbationII} These are, namely, the halogen diatomic and hydride molecules I$_2$, I$_2^-$, At$_2^-$, HI, HAt, HTs. We recall that the PT2$^\prime$ calculation differs from the full PT2 one by an uncoupled, rather than coupled, solution of Eq. (\ref{eqn:pert_2}). We refer to Part II for computational details on the molecular systems. 

The GKS eigenvalue spectra are provided in Figs. S1 and S2 of the ESI. Noticeable differences are only found on the single most difficult case of At$_2^-$, which, as previously noted, would ideally require an ensemble, rather than pure-state treatment.\cite{desmarais2021perturbationII} For At$_2^-$, in Fig. S2, the 1c-SCF and PT2$^\prime$ HOMO-LUMO gaps are 2.20 and 2.45 eV, while the PT2 gap is 2.04 eV and the 2c-SCF one is 2.00 eV. Another difficult case is represented by the 7$p$ superheavy-element Tennessine hydride HTs system of Fig. S1. In this case, the 1c-SCF, PT2$^\prime$, PT2 and 2c-SCF gaps are 7.82, 5.76, 5.66 and 6.12 eV. In all other cases, of Figs. S1 and S2, the PT2$^\prime$, PT2 and 2c-SCF eigenvalue spectra are essentially matching.

As for density variables of SCDFT, Fig. S3 reports contour maps for the I$_2^-$ molecule. Here small difference between PT2$^\prime$ and 2c-SCF are only noticeable on the particle-current density $\mathbf{j}$, and these differences are resolved with the full PT2 calculation.

\subsection{Electronic Band Structure}

We here validate the present PT treatment for calculation of GKS band structures. Predicted band gaps of the tungsten dichalcogenide series are reported in Table \ref{tab:band_gaps}, as well as differences w.r.t. 2c-SCF values through first and second orders in PT ($\vert \Delta_1 \vert$ and $\vert \Delta_{2^\prime} \vert$). Here the PT2$^\prime$ gaps are always found to improve upon PT1, with regards to the comparison against 2c-SCF. Errors larger than $0.01$ eV through second order are only found in the single most difficult case of $\text{WLv}_2$, where $\vert \Delta_{2^\prime} \vert=0.09$ eV. For $\text{WLv}_2$, the scalar-relativistic 1c-SCF gap is $E_g^\text{1c}=1.90$ eV, while the 2c-SCF one is $E_g^\text{2c}=0.82$ eV. In this particularly challenging case, first-order PT wrongly predicts a metallic system, with a gap of $0.00$ eV. Remarkably, the correct description is recovered  with the PT2$^\prime$ treatment, where the predicted gap becomes $E_g^{\text{PT2}^\prime}=0.91$ eV, in good agreement with the 2c-SCF value.

\begin{table}[b!]
\caption{Calculated energy band gaps and relative differences w.r.t. 2c-SCF (eV).}
\label{tab:band_gaps}
\vspace{5pt}
\begin{tabular}{lccccccccc}
\hline
\hline
&&\\
& & $E_g^\text{1c}$ & $E_g^\text{PT1}$ & $E_g^{\text{PT2}^\prime}$ & $E_g^\text{2c}$ & & $\vert \Delta_1 \vert$ & $\vert \Delta_{2^\prime} \vert$ \\
\hline
&&\\
\multirow{3}{*}{LC} & $\text{WSe}_2$ & $2.68$ & $2.31$ & $2.34$ & $2.33$ & & $0.02$ & $0.01$\\
& $\text{WTe}_2$ & $2.24$ & $1.80$ & $1.84$ & $1.83$ && $0.03$ &  $0.01$\\
& $\text{WPo}_2$ & $2.05$ & $1.35$ & $1.45$ & $1.44$ &&  $0.09$& $0.01$ \\
&&\\
\multirow{4}{*}{SC} & $\text{WSe}_2$ & $2.68$ & $2.30$ & $2.34$ & $2.33$ && $0.02$ & $0.01$\\
& $\text{WTe}_2$ & $2.08$ & $1.61$ & $1.66$ & $1.65$ && $0.04$ & $0.01$\\
& $\text{WPo}_2$ & $1.83$ & $1.08$ & $1.19$ & $1.18$ && $0.11$  & $0.01$\\
& $\text{WLv}_2$ & $1.90$ & $0.00$ & $0.91$ & $0.82$ && $0.82$ & $0.09$\\
&&\\
\hline 
\hline
\end{tabular}
\end{table}

The full electronic band structures of the tungsten dichalcogenide series, as calculated with the SC potentials, are provided in Fig. \ref{fig:bands}. Because these layers lack an inversion center, the scalar-relativistic bands (in the black dashed line) are spin-split by inclusion of SOC. As expected, the magnitude of the spin-splitting increases progressively along the series. In the case of the two lightest-element systems, being WSe$_2$ and WTe$_2$ in the upper panels, the PT2$^\prime$ bands (in magenta) are essentially superimposed on the 2c-SCF ones (in blue), and visible improvements over the PT1 treatment (in turquoise) can be noticed, especially for the virtual levels. Going down the series, to the bottom panels, for WPo$_2$ and WLv$_2$, where SOC is enhanced, differences between PT1 and PT2$^\prime$/2c-SCF become more evident. In the case of WPo$_2$, small differences between PT2$^\prime$ and 2c-SCF begin to appear, and are again more apparent for virtual bands. In the single most challenging case of WLv$_2$, first-order PT appears to have completely failed, as the turquoise PT1 bands display very large differences w.r.t. 2c-SCF. Here, for WLv$_2$, the improvement provided by the approximate second-order corrections is remarkable, as the magenta PT2$^\prime$ bands show instead a good agreement with the blue 2c-SCF ones. The occupied WLv$_2$ PT2$^\prime$ bands nearly match the 2c-SCF ones, but visible differences remain on virtual bands.

\begin{widetext}
\begin{center}
\begin{figure}[h!]
\centering
\includegraphics[width=18cm]{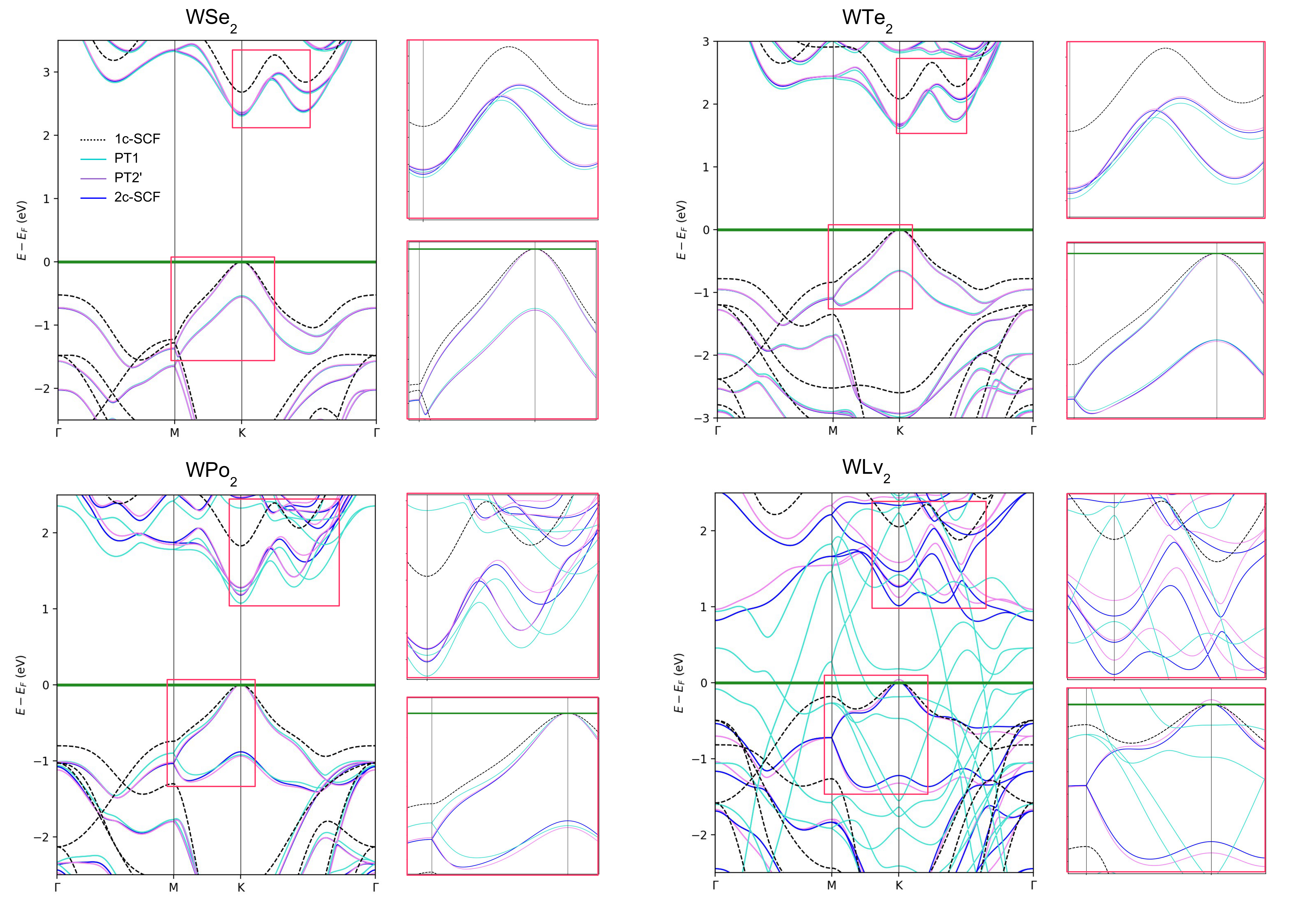}
\caption{Electronic band structures of the tungsten dichalcogenide 2D systems, at the scalar-relativistic (black dashed lines) level, and including SOC, both through first- or approximate second-order PT (PT1 in turquoise and PT2$^\prime$ in magenta), or the reference 2c-SCF (in blue). Close-up views on the occupied and virtual levels are also provided for each system.}
\label{fig:bands}
\end{figure}
\end{center}
\end{widetext}

\subsection{Density Variables of Spin-Current DFT}

In the following, we report on calculations that quantify the accuracy of the CPKS approach to reproduce the effect of SOC on the two-component complex density matrix, through density variables of SCDFT: i) the electron density $\rho$, and ii) the spin-current density $\mathbf{J}^z$. The density variables are for calculations employing the LC, rather than SC, potential so as to present results that are more strongly influenced by contributions originating from valence orbitals (therefore for WSe$_2$, WTe$_2$ and WPo$_2$).

\subsubsection{The Electron Density}

Fig. \ref{fig:eldens} reports on calculations of the effect of SOC on the electron density in WSe$_2$, WTe$_2$ and WPo$_2$, as determined by different approaches (2c-SCF, and CPKS).  In first order, the PT1 density $\rho_\textup{PT1}$ coincides with the scalar-relativistic density $\rho_\textup{1c}$, because the real part of diagonal blocks of the first-order perturbed density are vanishing, from Eq. (\ref{eqn:rp1_0}). On the other hand, these same blocks of the perturbed density do not vanish, with the PT2$^\prime$ approach, so that the $\rho_{\textup{PT2}^\prime}$ provides a description of the effect of SOC on the electron density. 

We report the quantity $\Delta\rho_\textup{SOC} = \rho_\textup{2c} - \rho_\textup{1c}$ providing the total effect of SOC on $\rho$, as well as the difference of CPKS and 2c-SCF values $\Delta\rho_{\textup{PT2}^\prime-2c} = \rho_{\textup{PT2}^\prime} -\rho_\textup{2c}$. The figure provides electron density data on two distinct planes, being i) the plane containing the chalcogen atoms (upper panels) and ii) the plane containing the tungsten atoms (bottom panels). Generally speaking, the plots show that $\Delta\rho_{\textup{PT2}^\prime-2c}$ is always smaller than $\Delta\rho_\textup{SOC}$, and $\Delta \rho_\textup{SOC}$ as well as $\Delta\rho_{\textup{PT2}^\prime-2c}$ are always largest on the plane containing the tunsten atoms. The maxima of $\Delta\rho_{\textup{PT2}^\prime-2c}$ is smaller than the one for  $\Delta \rho_\textup{SOC}$, by a factor of about 3 (chalcogen plane for WSe$_2$, tunsten plane for WTe$_2$ and WPo$_2$) to about 9 (chalcogen plane for WTe$_2$ and WPo$_2$) with larger values of $\Delta\rho_{\textup{PT2}^\prime-2c}$ in the core, rather than bonding, region of the atoms. To further quantify the differences, density profiles in WTe$_2$ are provided along the W-W and W-Te directions in bottom right panels of Fig. \ref{fig:eldens}. The profiles confirm that $\Delta\rho_{\textup{PT2}^\prime-2c}$ has a peak close to the core of the W atom (at distance of about 0.5 \AA{} from the W nucleus) and smaller values in the bonding regions.

\begin{widetext}
\begin{center}
\centering
\begin{figure}[h!]
\includegraphics[width=16cm]{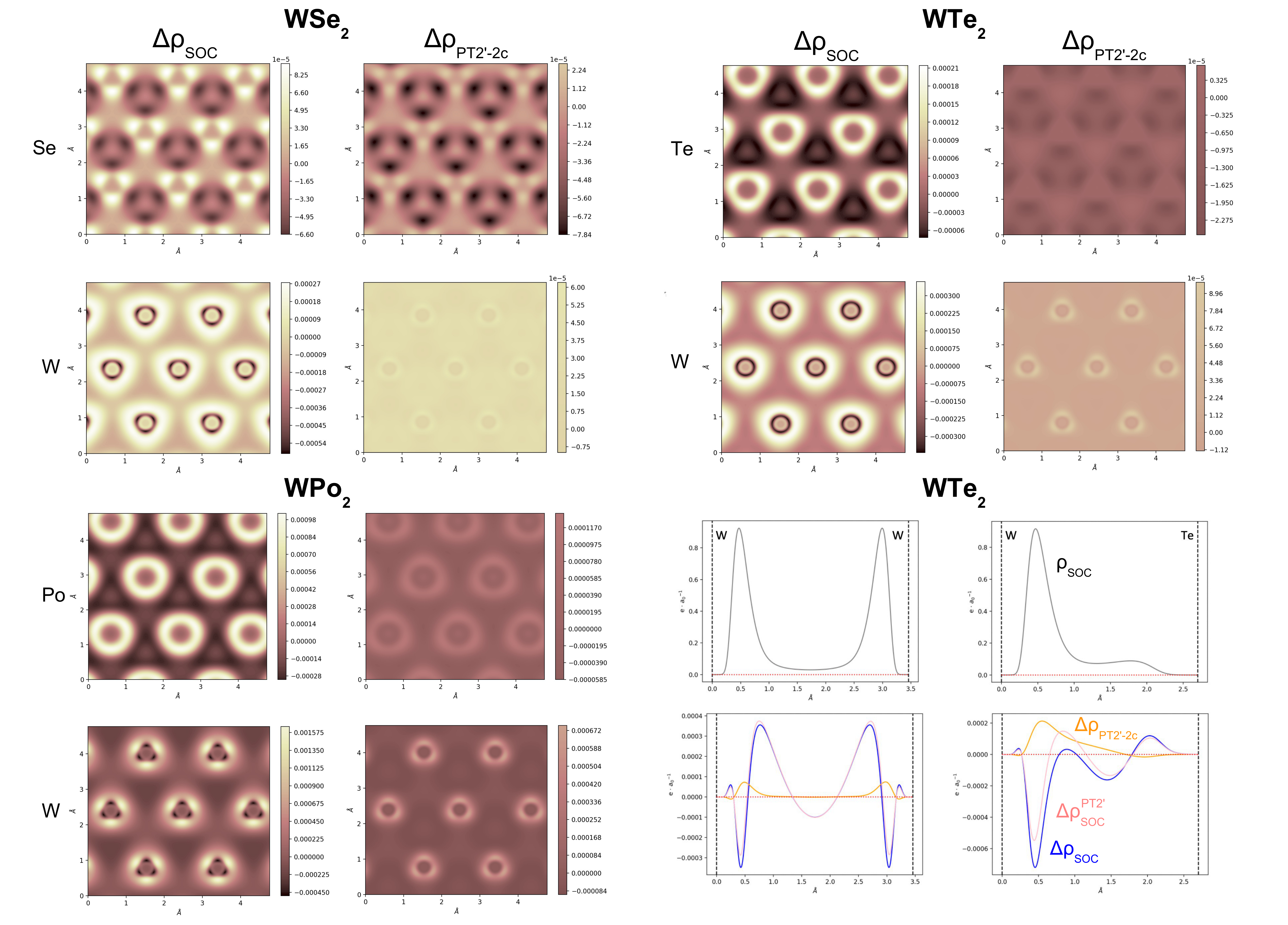}
\caption{Effect of SOC on the electron density $\rho$ of WSe$_2$, WTe$_2$ and WPo$_2$. For each of the three systems, 2D maps are reported on two planes: the plane of the chalcogen atoms (top panels) and the plane of the W atoms (bottom panels). For each system and each plane, two quantities are mapped: the effect of SOC on the electron density $\Delta\rho_\textup{SOC} = \rho_\textup{2c} - \rho_\textup{1c}$ (left panels) and the difference between the PT2$^\prime$ density and the reference $2c$ one $\Delta\rho_{\textup{PT2}^\prime-2c} = \rho_{\textup{PT2}^\prime} -\rho_\textup{2c}$ (right panels). The bottom right panels show electron density profiles of WTe$_2$ along W-W (left) and W-Te (right) directions: upper panels report the profile of the total density $\rho_\textup{SOC}$ obtained from the reference $2c$ calculation while bottom panels report profiles of $\Delta\rho_\textup{SOC}$ (blue) and $\Delta\rho_{\textup{PT2}^\prime-2c}$ (yellow) defined above, along with profiles of $\Delta\rho_\textup{SOC}^{\textup{PT2}^\prime} = \rho_{\textup{PT2}^\prime} - \rho_\textup{1c}$ (pink).}
\label{fig:eldens}
\end{figure}
\end{center}
\end{widetext}

\subsubsection{The Spin Current Density}

We now discuss the accuracy of the PT approach to reproduce spin-current densities of the tungsten dichalcogenide series. In the case of $\mathbf{J}^z=\mathbf{J}^\uparrow-\mathbf{J}^\downarrow$ the $z$-component spin-current density represents the local velocity field for transport of spin-magnetization $m_z$ (and similar for $\mathbf{J}^x$ and $\mathbf{J}^y$ being the velocity fields for transport of $m_x$ and $m_y$). Our calculations reproduce all three $\mathbf{J}^x$, $\mathbf{J}^y$ and $\mathbf{J}^z$ with similar accuracy, and we therefore report on results of only $\mathbf{J}^z$ for sake of brevity. 

The results of the calculations on WSe$_2$, WTe$_2$ and WPo$_2$ with the LC potentials are reported in Fig. \ref{fig:spincur}. At variance with the electron density $\rho$, the spin-currents $\mathbf{J}^x$, $\mathbf{J}^y$ and $\mathbf{J}^z$ are not vanishing in first-order, but are vanishing for scalar-relativistic calculations (i.e. in zeroth-order), and we correspondingly report three maps (total 2c-SCF $\mathbf{J}^z$ on the left panels, $\Delta\mathbf{J}^z_{\textup{PT1}-2c} = \mathbf{J}^z_{\textup{PT1}} -\mathbf{J}^z_\textup{2c}$ on the middle panels and $\Delta\mathbf{J}^z_{\textup{PT2$^\prime$}-2c} = \mathbf{J}^z_{\textup{PT2$^\prime$}} -\mathbf{J}^z_\textup{2c}$ on the right panels). The maps of $\Delta\mathbf{J}^z_{\textup{PT1}-2c}$ and $\Delta\mathbf{J}^z_{\textup{PT2$^\prime$}-2c}$ share a common color scale. With $\mathbf{J}^z$ being a vector field, here the color intensities represent the vector magnitude $\vert \mathbf{J}^z \vert$, while the arrow lengths and directions reflect the vector projection of $\mathbf{J}^z$ onto the planes that contain the chalcogen atoms (upper panels) and the tungsten atoms (lower panels). The $z$ Cartesian direction is perpendicular to the plane, and the $\mathbf{J}^z$ vectors are found in the figures to form circular orbits in the planes. The arrows are not visible in the difference maps, indicating that the CPKS approach provides very accurate orientations for the spin-current densities. Colors, on the other hand are visible, but are about two (in the case of  WSe$_2$ and WTe$_2$) to about one (in the case of WPo$_2$) orders of magnitude smaller than the total $\mathbf{J}^z$, indicating that also the vector magnitude $\vert \mathbf{J}^z \vert$ is well reproduced by all CPKS calculations. The $\textup{PT2}^\prime$ treatment usually improves over the $\textup{PT1}$ treatment, as the maps of $\Delta\mathbf{J}^z_{\textup{PT2$^\prime$}-2c}$ usually have color intensities that are smaller than those of $\Delta\mathbf{J}^z_{\textup{PT1}-2c}$. The only exception is for the map of $\mathbf{J}^z$ in the plane containing the tungsten atoms for the lightest element system WSe$_2$, where more intense colors are visible in the core region of tungsten atoms. In this case (W-plane of WSe$_2$) even though the $\textup{PT2}^\prime$ treatment worsens the description of spin-current densities in inner shells, the differences remain very small (around two orders of magnitude smaller than the total $\mathbf{J}^z$), and the agreement is improved for the description of $\mathbf{J}^z$ in the valence region, because darker colors are observed in the region separating the tungsten atoms.

\begin{widetext}
\begin{center}
\centering
\begin{figure}[h!]
\includegraphics[width=18cm]{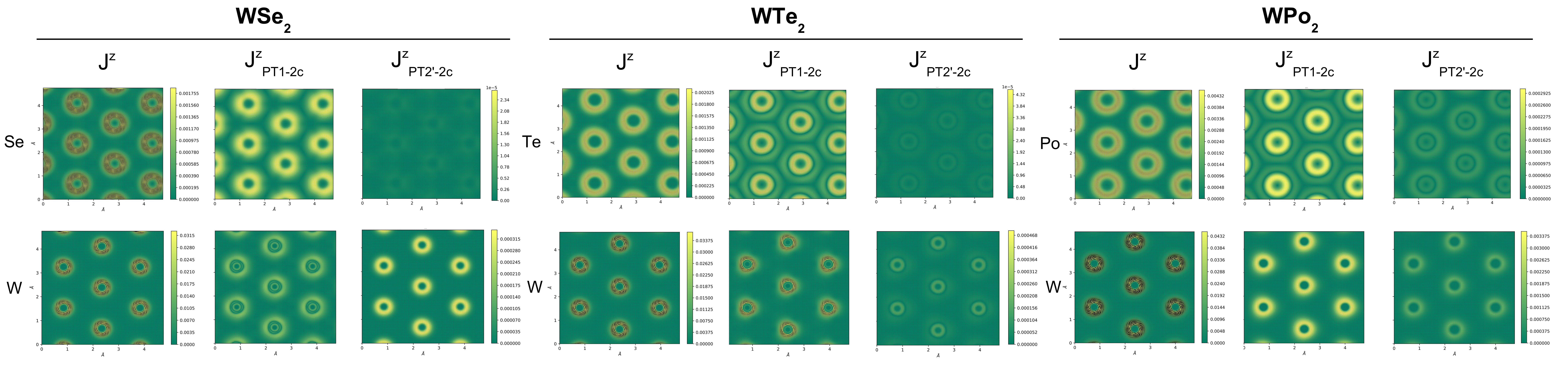}
\caption{Effect of SOC on the spin-current density $\mathbf{J}^z$ of WSe$_2$, WTe$_2$ and WPo$_2$. The color intensities represent the magnitude $\vert \mathbf{J}^z \vert$, while the arrow lengths and directions reflect the vector projection of $\mathbf{J}^z$ onto the plane. For each system and each plane, three quantities are mapped: the effect of SOC on the spin-current density $\mathbf{J}^z$ (left panels), the difference between the PT1 spin-current density and the reference $2c$ one $\Delta\mathbf{J}^z_{\textup{PT1}-2c} = \mathbf{J}^z_{\textup{PT1}} -\mathbf{J}^z_\textup{2c}$ (middle panels), and finally  $\Delta\mathbf{J}^z_{\textup{PT2$^\prime$}-2c} = \mathbf{J}^z_{\textup{PT2$^\prime$}} -\mathbf{J}^z_\textup{2c}$ (right panels).}
\label{fig:spincur}
\end{figure}
\end{center}
\end{widetext}

\section{Conclusions}

A previously proposed molecular non-canonical coupled-perturbed Kohn-Sham density functional theory (KS-DFT)/Hartree-Fock (HF) treatment for spin-orbit coupling has been generalized to infinite periodic systems. Explicit expressions have been provided for the total energy through 3rd-order, which satisfy the 2N + 1 rule (i.e. requiring only the 1st-order perturbed wave function for its computation). Satisfaction of the 2N + 1 rule has been achieved for periodic systems by partitioning the key matrices in terms of components that transform with even and odd parity upon inversion in reciprocal space ($\mathbf{k} \to \mathbf{-k}$). Second-order corrections to the perturbed wave function (and one-electron properties) are calculated at the uncoupled-perturbed level of theory (and this approximation has been justified on the well-characterized diatomic halogen series of molecules). 

The perturbation-theory approach has been validated for calculating the total energy, electronic band structure and density variables of spin-current DFT on the tungsten dichalcogenide hexagonal 2D series (i.e. WSe$_2$, WTe$_2$, WPo$_2$ and WLv$_2$), including challenging 6p and 7p elements. The computed properties through second- or third-order match well with those from reference two-component self-consistent field (2c-SCF) calculations. For total energies, $E^{(3)}$ is found to consistently improve the agreement against the 2c-SCF values. For electronic band structures, visible differences w.r.t. 2c-SCF remain through second-order in only the single-most difficult case of WLv$_2$. As for density variables of spin-current DFT, the perturbed electron density, being vanishing in first-order, is the most challenging for the perturbation theory. Visible differences in the electron densities are, however, found to be largest close to the core region of atoms and smaller in the valence region relevant to chemical bonding. Perturbed spin-current densities, on the other hand, are well-reproduced in all tested cases. Our coupled-perturbed approach thus provides an accurate alternative to 2c-SCF, with potential for over an order of magnitude savings in computation times. Moreover, it constitutes a convenient starting point for future improvements to treat multi-reference systems, where an ensemble treatment would be necessary.

\newpage

\appendix
\section{Virtual-Occupied Blocks of the Second-Order Matrix of Orbital Rotations}
\label{app:U2vo}
The simultaneous non-canonical solution of Eqs. (\ref{eqn:ortho2}) and (\ref{eqn:pert_2}) yields, for the virtual-occupied (VO) blocks:
\begin{widetext}
\begin{subequations}
\begin{eqnarray}
\label{eqn:UvoR2}
\mathbf{U}^{\sigma^\prime \sigma \mathcal{R} (2)}_{VO} \left( \mathbf{k} \right) &=& \frac{1}{ \epsilon_V^{\sigma^\prime(0)} \left( \mathbf{k} \right)  - \epsilon_O^{\sigma(0)} \left( \mathbf{k} \right)  } \Bigg\{ \sum_{\mathcal{C}} \sum_{\sigma^{\prime \prime}} \Bigg[ \sum_{O^\prime} \mathbf{U}^{\sigma^\prime \sigma^{\prime \prime} \mathcal{C} (1)}_{VO^\prime} \left( \mathbf{k} \right)  \mathbf{G}^{\sigma^{\prime \prime} \sigma \mathcal{C} (1)}_{O^\prime O} \left( \mathbf{k} \right) \nonumber \\
&-& \sum_{V^\prime} \mathbf{G}^{\sigma^\prime \sigma^{\prime \prime} \mathcal{C} (1)}_{V V^\prime} \left( \mathbf{k} \right) \mathbf{U}^{\sigma^{\prime \prime} \sigma \mathcal{C} (1)}_{V^\prime O} \left( \mathbf{k} \right)  \Bigg] - \mathbf{G}^{\sigma^{\prime} \sigma \mathcal{R} (2)}_{V O} \left( \mathbf{k} \right)  \Bigg\} = - \left[ \mathbf{U}^{\sigma \sigma^\prime \mathcal{R} (2)}_{OV} \left( \mathbf{k} \right) \right]^\ast
\end{eqnarray}
and:
\begin{eqnarray}
\label{eqn:UvoI2}
\mathbf{U}^{\sigma^\prime \sigma \mathcal{I} (2)}_{VO} \left( \mathbf{k} \right) &=& \frac{1}{ \epsilon_V^{\sigma^\prime(0)} \left( \mathbf{k} \right)  - \epsilon_O^{\sigma(0)} \left( \mathbf{k} \right)   } \Bigg\{ \sum_{\mathcal{C}} \sum_{\mathcal{C}^\prime \ne \mathcal{C}}\sum_{\sigma^{\prime \prime}} \Bigg[ \sum_{O^\prime} \mathbf{U}^{\sigma^\prime \sigma^{\prime \prime} \mathcal{C} (1)}_{VO^\prime} \left( \mathbf{k} \right)  \mathbf{G}^{\sigma^{\prime \prime} \sigma \mathcal{C}^\prime (1)}_{O^\prime O} \left( \mathbf{k} \right) \nonumber \\
&-& \sum_{V^\prime} \mathbf{G}^{\sigma^\prime \sigma^{\prime \prime} \mathcal{C} (1)}_{V V^\prime} \left( \mathbf{k} \right) \mathbf{U}^{\sigma^{\prime \prime} \sigma \mathcal{C}^\prime (1)}_{V^\prime O} \left( \mathbf{k} \right)  \Bigg] - \mathbf{G}^{\sigma^{\prime} \sigma \mathcal{I} (2)}_{V O} \left( \mathbf{k} \right)  \Bigg\} = - \left[ \mathbf{U}^{\sigma \sigma^\prime \mathcal{I} (2)}_{OV} \left( \mathbf{k} \right) \right]^\ast
\end{eqnarray}
\end{subequations}
\end{widetext}

To obtain Eqs. (\ref{eqn:UvoR2}) and (\ref{eqn:UvoI2}), we insert Eqs. (\ref{eqn:eps_VO})-(\ref{eqn:eps_OO}) into Eq. (\ref{eqn:pert_2}).

\section{Lagrange-Multipliers in Second Order}
\label{app:E2}
Inserting Eqs. (\ref{eqn:U1C_VO})  (\ref{eqn:UooR2}), (\ref{eqn:UooI2}),  (\ref{eqn:UvoR2}) and (\ref{eqn:UvoI2}) into Eq. (\ref{eqn:pert_2}) provides the following set of non-canonical second-order Lagrange multipliers:
\begin{subequations}
\begin{eqnarray}
\label{eqn:eps2_oor}
\mathbf{E}_{OO^\prime}^{\sigma \sigma^\prime \mathcal{R} (2)} (\mathbf{k}) = \mathbf{G}_{OO^\prime}^{\sigma \sigma^{\prime}\mathcal{R} (2)}(\mathbf{k}) +  \sum_{\mathcal{C}} \sum_{\sigma^{\prime \prime}} \sum_{V}  \nonumber \\
\times \Big[  \frac{1}{2} \left( \epsilon_{O}^{\sigma (0)} (\mathbf{k}) - \epsilon_{O^\prime}^{\sigma^{\prime} (0)} (\mathbf{k}) \right)  \mathbf{U}^{\sigma \sigma^{\prime \prime} \mathcal{C} (1)}_{OV} (\mathbf{k}) \mathbf{U}^{\sigma^{\prime \prime} \sigma^{\prime} \mathcal{C}(1)}_{VO^\prime} (\mathbf{k}) \nonumber \\+ \left( \epsilon_{V}^{\sigma^{\prime \prime} (0)} (\mathbf{k}) - \epsilon_{O}^{\sigma (0)} (\mathbf{k}) \right) \mathbf{U}_{OV}^{\sigma \sigma^{\prime \prime} \mathcal{C} (1)} (\mathbf{k}) \mathbf{U}_{V O^\prime}^{\sigma^{\prime \prime} \sigma^\prime \mathcal{C}(1)}(\mathbf{k}) \Big] \;,\end{eqnarray}and:
\begin{eqnarray}
\label{eqn:eps2_ooi}
\mathbf{E}_{OO^\prime}^{\sigma \sigma^\prime \mathcal{I} (2)} (\mathbf{k}) = \mathbf{G}_{OO^\prime}^{\sigma \sigma^{\prime}\mathcal{I} (2)} (\mathbf{k}) +   \sum_{\mathcal{C}} \sum_{\mathcal{C}^\prime \ne \mathcal{C}}  \sum_{\sigma^{\prime \prime}} \sum_{V}   \nonumber \\
\times \Big[  \frac{1}{2} \left( \epsilon_{O}^{\sigma (0)}(\mathbf{k}) - \epsilon_{O^\prime}^{\sigma^{\prime} (0)} (\mathbf{k})\right)  \mathbf{U}^{\sigma \sigma^{\prime \prime} \mathcal{C} (1)}_{OV} (\mathbf{k})  \mathbf{U}^{\sigma^{\prime \prime} \sigma^{\prime} \mathcal{C}^\prime (1)}_{VO^\prime} (\mathbf{k}) \nonumber \\
+ \left( \epsilon_{V}^{\sigma^{\prime \prime} (0)} (\mathbf{k}) - \epsilon_{O}^{\sigma (0)} (\mathbf{k}) \right) \mathbf{U}_{OV}^{\sigma \sigma^{\prime \prime} \mathcal{C} (1)} (\mathbf{k}) \mathbf{U}_{V O^\prime}^{\sigma^{\prime \prime} \sigma^\prime \mathcal{C}^\prime (1)} (\mathbf{k}) \Big] \;,
\end{eqnarray}
\end{subequations}
where we have introduced:
\begin{equation}
\mathbf{E}_{OO^\prime}^{\sigma \sigma^\prime (2)} (\mathbf{k}) = \mathbf{E}_{OO^\prime}^{\sigma \sigma^\prime \mathcal{R} (2)} (\mathbf{k}) + \mathbf{E}_{OO^\prime}^{\sigma \sigma^\prime \mathcal{I} (2)} (\mathbf{k})
\end{equation}
Exactly analogous expressions can also be worked out for $\mathbf{E}_{VV^\prime}^{\sigma \sigma^\prime (2)} (\mathbf{k})$. 

\section{Direct-Space Perturbed Density Matrix in First and Second Orders}
\label{app:pg}
Inserting Eq. (\ref{eqn:P1}) into Eq. (\ref{eqn:p1_g1}) provides:
\begin{widetext}
\begin{eqnarray}
\label{eqn:p1_g2}
\mathbf{P}^{\sigma \sigma^\prime(1)} ( \mathbf{g} ) &=&  \frac{1}{\Omega} \int^\prime d \mathbf{k} \ \Big\{ e^{\imath \mathbf{k} \cdot \mathbf{g}} \mathbf{C}^{\sigma \sigma(0)} (\mathbf{k}) \mathbf{f}_\sigma (\mathbf{k}) \left[ \mathbf{U}^{\sigma \sigma^\prime(1)} (\mathbf{k}) \right]^\dagger \left[ \mathbf{C}^{\sigma^\prime \sigma^\prime (0)} (\mathbf{k}) \right]^\dagger \nonumber \\
&+& e^{-\imath \mathbf{k} \cdot \mathbf{g}} \mathbf{C}^{\sigma \sigma(0)} (-\mathbf{k}) \mathbf{f}_\sigma (-\mathbf{k}) \left[ \mathbf{U}^{\sigma \sigma^\prime(1)} (-\mathbf{k}) \right]^\dagger \left[ \mathbf{C}^{\sigma^\prime \sigma^\prime (0)} (-\mathbf{k}) \right]^\dagger \nonumber \\
&+& e^{\imath \mathbf{k} \cdot \mathbf{g}} \mathbf{C}^{\sigma \sigma(0)} (\mathbf{k}) \mathbf{U}^{\sigma \sigma^\prime(1)} (\mathbf{k}) \mathbf{f}_{\sigma^\prime} (\mathbf{k}) \left[ \mathbf{C}^{\sigma^\prime \sigma^\prime (0)} (\mathbf{k}) \right]^\dagger + e^{-\imath \mathbf{k} \cdot \mathbf{g}} \mathbf{C}^{\sigma \sigma(0)} (-\mathbf{k}) \mathbf{U}^{\sigma \sigma^\prime(1)} (-\mathbf{k}) \mathbf{f}_{\sigma^\prime} (-\mathbf{k}) \left[ \mathbf{C}^{\sigma^\prime \sigma^\prime (0)} (-\mathbf{k}) \right]^\dagger \Big\} 
\end{eqnarray}
Now, to relate elements at $\mathbf{k}$ and $-\mathbf{k}$, we also require:
\begin{equation}
\label{eqn:f_k_mk}
\mathbf{f}_{\sigma} (\mathbf{k}) = \mathbf{f}_{\sigma} (-\mathbf{k})
\end{equation}
which follows directly from Eq. (\ref{eqn:e_k_mk}) since, in pure-state GKS-DFT calculations, degenerate bands must have the same occupation. Then, inserting Eqs. (\ref{eqn:c_k_mk}), (\ref{eqn:e_k_mk}) and (\ref{eqn:UN_k_mk}) into Eq. (\ref{eqn:p1_g2}) gives, after combining complex-conjugates:
\begin{eqnarray}
\label{eqn:p1_g}
\mathbf{P}^{\sigma \sigma^\prime(1)} ( \mathbf{g} ) &=& \frac{2}{\Omega} \int^\prime d \mathbf{k} \ \Big\{ \Re \left( e^{\imath \mathbf{k} \cdot \mathbf{g}} \mathbf{C}^{\sigma \sigma(0)}_O (\mathbf{k}) \mathbf{f}_{\sigma O} (\mathbf{k}) \left[ \mathbf{U}^{\sigma \sigma^\prime \mathcal{R} (1)}_{OV} (\mathbf{k}) \right]^\dagger \left[ \mathbf{C}^{\sigma^\prime \sigma^\prime (0)}_V (\mathbf{k}) \right]^\dagger \right) \ \nonumber \\
&+& \imath \Im \left( e^{\imath \mathbf{k} \cdot \mathbf{g}} \ \mathbf{C}^{\sigma \sigma(0)}_O (\mathbf{k}) \mathbf{f}_{\sigma O} (\mathbf{k}) \left[ \mathbf{U}^{\sigma \sigma^\prime \mathcal{I}(1)}_{OV} (\mathbf{k}) \right]^\dagger \left[ \mathbf{C}^{\sigma^\prime \sigma^\prime (0)}_V (\mathbf{k}) \right]^\dagger \right) \ \nonumber \\
&+& \Re \left( e^{\imath \mathbf{k} \cdot \mathbf{g}} \mathbf{C}^{\sigma \sigma(0)}_V (\mathbf{k}) \mathbf{U}^{\sigma \sigma^\prime \mathcal{R} (1)}_{VO} (\mathbf{k}) \mathbf{f}_{\sigma^\prime O} (\mathbf{k}) \ \left[ \mathbf{C}^{\sigma^\prime \sigma^\prime (0)}_O (\mathbf{k}) \right]^\dagger \right) + \i \Im \left( e^{\imath \mathbf{k} \cdot \mathbf{g}} \mathbf{C}^{\sigma \sigma(0)}_V (\mathbf{k}) \mathbf{U}^{\sigma \sigma^\prime\mathcal{I} (1)}_{VO} (\mathbf{k}) \mathbf{f}_{\sigma^\prime O} (\mathbf{k}) \left[ \mathbf{C}^{\sigma^\prime \sigma^\prime (0)}_O (\mathbf{k}) \right]^\dagger \right) \Big\} 
\end{eqnarray}
In second order, an expansion of the CO coefficients of Eq. (\ref{eqn:P}) using Eq. (\ref{eqn:CUC}) provides:
\begin{eqnarray}
\label{eqn:p2_k}
\mathbf{P}^{\sigma \sigma^\prime(2)} ( \mathbf{k} ) &=& \mathbf{C}^{\sigma \sigma(0)} (\mathbf{k}) \mathbf{f}_\sigma (\mathbf{k}) \ \left[ \mathbf{U}^{\sigma \sigma^\prime(2)} (\mathbf{k}) \right]^\dagger \left[ \mathbf{C}^{\sigma^\prime \sigma^\prime (0)} (\mathbf{k}) \right]^\dagger + \mathbf{C}^{\sigma \sigma(0)} (\mathbf{k}) \mathbf{U}^{\sigma \sigma^\prime(2)} (\mathbf{k}) \mathbf{f}_{\sigma^\prime} (\mathbf{k}) \left[ \mathbf{C}^{\sigma^\prime \sigma^\prime (0)} (\mathbf{k}) \right]^\dagger \ \nonumber \\
&+& \sum_{\sigma^{\prime \prime}} \mathbf{C}^{\sigma \sigma(0)} (\mathbf{k}) \mathbf{U}^{\sigma \sigma^{\prime \prime}(1)} (\mathbf{k}) \mathbf{f}_{\sigma^{\prime \prime}} (\mathbf{k}) \left[ \mathbf{U}^{\sigma^{\prime \prime} \sigma^\prime(1)} (\mathbf{k}) \right]^\dagger \left[ \mathbf{C}^{\sigma^\prime \sigma^\prime (0)} (\mathbf{k}) \right]^\dagger 
\end{eqnarray}

Then, proceeding as in Eqs. (\ref{eqn:p1_g2}) and (\ref{eqn:p1_g}), using Eqs. (\ref{eqn:UooR2}),  (\ref{eqn:UooI2}), as well as Eqs. (\ref{eqn:UvoR2}) and (\ref{eqn:UvoI2}), provides:
\begin{eqnarray}
\label{eqn:p2_g}
\mathbf{P}^{\sigma \sigma^\prime(2)} ( \mathbf{g} ) &=&  \frac{2}{\Omega} \int^\prime d \mathbf{k} \Big\{ \Re \left( e^{\imath \mathbf{k} \cdot \mathbf{g}} \mathbf{C}^{\sigma \sigma(0)} (\mathbf{k}) \mathbf{f}_\sigma (\mathbf{k}) \left[ \mathbf{U}^{\sigma \sigma^\prime \mathcal{R} (2)} (\mathbf{k}) \right]^\dagger \left[ \mathbf{C}^{\sigma^\prime \sigma^\prime (0)} (\mathbf{k}) \right]^\dagger \right) \nonumber \\
&+& \imath \Im \left( e^{\imath \mathbf{k} \cdot \mathbf{g}} \mathbf{C}^{\sigma \sigma(0)} (\mathbf{k}) \mathbf{f}_\sigma (\mathbf{k}) \left[ \mathbf{U}^{\sigma \sigma^\prime \mathcal{I} (2)} (\mathbf{k}) \right]^\dagger \left[ \mathbf{C}^{\sigma^\prime \sigma^\prime (0)} (\mathbf{k}) \right]^\dagger \right) \ \nonumber \\
&+& \Re \left( e^{\imath \mathbf{k} \cdot \mathbf{g}} \mathbf{C}^{\sigma \sigma(0)} (\mathbf{k}) \mathbf{U}^{\sigma \sigma^\prime \mathcal{R} (2)} (\mathbf{k}) \mathbf{f}_{\sigma^\prime} (\mathbf{k}) \left[ \mathbf{C}^{\sigma^\prime \sigma^\prime (0)} (\mathbf{k}) \right]^\dagger \right) + \imath \Im \left( e^{\imath \mathbf{k} \cdot \mathbf{g}} \mathbf{C}^{\sigma \sigma(0)} (\mathbf{k}) \mathbf{U}^{\sigma \sigma^\prime \mathcal{I} (2)} (\mathbf{k}) \mathbf{f}_{\sigma^\prime} (\mathbf{k}) \left[ \mathbf{C}^{\sigma^\prime \sigma^\prime (0)} (\mathbf{k}) \right]^\dagger \right) \nonumber \\
&+& \sum_{\mathcal{C}} \sum_{\sigma^{\prime \prime}} \Re \left( e^{\imath \mathbf{k} \cdot \mathbf{g}} \mathbf{C}^{\sigma \sigma(0)} (\mathbf{k}) \mathbf{U}^{\sigma \sigma^{\prime \prime} \mathcal{C}(1)} (\mathbf{k}) \mathbf{f}_{\sigma^{\prime \prime}} (\mathbf{k}) \left[ \mathbf{U}^{\sigma^{\prime \prime} \sigma^\prime \mathcal{C} (1)} (\mathbf{k}) \right]^\dagger \left[ \mathbf{C}^{\sigma^\prime \sigma^\prime (0)} (\mathbf{k}) \right]^\dagger \right) \nonumber \\
&+& \sum_{\mathcal{C}} \sum_{\mathcal{C}^\prime \ne \mathcal{C}}  \sum_{\sigma^{\prime \prime}} \imath \Im \left( e^{\imath \mathbf{k} \cdot \mathbf{g}} \mathbf{C}^{\sigma \sigma(0)} (\mathbf{k}) \mathbf{U}^{\sigma \sigma^{\prime \prime} \mathcal{C}(1)} (\mathbf{k}) \mathbf{f}_{\sigma^{\prime \prime}} (\mathbf{k}) \left[ \mathbf{U}^{\sigma^{\prime \prime} \sigma^\prime \mathcal{C}^\prime (1)} (\mathbf{k}) \right]^\dagger \left[ \mathbf{C}^{\sigma^\prime \sigma^\prime (0)} (\mathbf{k}) \right]^\dagger \right) \Big\}
\end{eqnarray}
\end{widetext}


\end{document}